\documentstyle[prl,aps,epsfig]{revtex}
\draft

\begin{document}
\twocolumn[
\hsize\textwidth\columnwidth\hsize\csname@twocolumnfalse\endcsname

\title{Metal-Kondo insulating transitions and transport in one dimension}
\author{Karyn Le Hur}
\address
{Theoretische Physik, ETH-H\"onggerberg, CH-8093 Z\"urich, Switzerland}
\maketitle

\begin{abstract}

We study two different metal-insulating transitions possibly
occurring in one-dimensional Kondo lattices.
First, we show how doping
the pure Kondo lattice model in the {\it strong}-coupling limit, results in a
Pokrovsky-Talapov transition. This produces a
conducting state with a
charge susceptibility diverging as the inverse of the doping, that seems
in agreement with numerical datas. Second, in the
{\it weak}-coupling region, Kondo insulating
transitions arise due to the consequent
renormalization of the backward Kondo scattering. Here, the interplay
between Kondo effect and electron-electron interactions gives rise to
significant phenomena in transport, in the high-temperature delocalized
(ballistic) regime. For repulsive interactions,
as a perfect signature of Kondo localization,
the 
conductivity is found to decrease monotonically with temperature.
When interactions become attractive, 
spin fluctuations in the electron (Luttinger-type) liquid are suddenly
lowered. The latter is {\it less} localized by magnetic impurities 
than for the repulsive counterpart, and as a result 
a large jump in the Drude weight and a maximum in the conductivity
arise in the entrance of the Kondo insulating phase. These can be
viewed as remnants of s-wave
superconductivity arising for attractive enough interactions. 
Comparisons with transport in the single impurity model are also
performed. We finally discuss the case of {\it randomly} distributed magnetic
defects, and the applications on persistent
currents of mesoscopic rings.

\end{abstract}

\vfill
 \twocolumn
\vskip.5pc ]
\narrowtext


\section{Introduction}

The interplay between magnetic impurities and itinerant electrons gives
rise to fascinating situations, in link with the so called Kondo 
effect\cite{Hewson-book}. Although the physics of one impurity 
in a metal is well understood\cite{Wilson}, 
when interactions are included or the number
of impurities increased, the problem remains largely open, solutions existing
however 
through Gutzwiller variational approximations\cite{Gutzwiller},
mean-field methods\cite{Bernard}, 
or still infinite-dimension treatments\cite{Georges-Kotliar}.

The one-dimensional (1D) models are usually much easier to handle than their
counterpart in higher dimensions. 
They can even prove to be exactly solvable, as
is the case for the Kondo model in 
a free electron gas\cite{Andrei}, or still the
1D Hubbard model\cite{Lieb-Wu}. 
Even for more complicated models, very powerful techniques,
such as bosonization or renormalization calculations, are still applicable
and generally give the correct physical results: For instance, 
these have allowed
to predict the generic Luttinger liquid concept\cite{Haldane,Schulz1}, induced
at low energy by weak electron-electron interactions. Such one-dimensional
physics
has received a 
consequent attention recently, due to for examples, advances in 
nanofabrication\cite{Tarucha}, the existence
of edge states in the fractional quantum Hall effect\cite{Tsui}
and the discovery of novel 1D materials such as 
carbon nanotubes\cite{Iijima}. Finally,
low dimensional
models can still provide valuable information on the role of correlation
effects in higher dimensions, e.g., on the physics of correlated fermions
in two dimensions (in link with high-$T_c$ materials) or, still
in our context,
on the phase diagram of the two-impurity
Kondo model in a three dimensional Landau-Fermi liquid\cite{Jones}.

Since the discovery of Anderson localization, non-magnetic
impurity effects in these Luttinger liquids (LL's) 
have always been a fascinating subject. The
two extreme situations, respectively of one or two impurities and
of a finite density of scatterers is now quite well understood\cite{Helene}. 
\vskip 0.05cm
In this paper, we
rather ponder the role of {\it magnetic} impurities in
the transport of LL's, using bosonization techniques. 
Precisely, we start with a conduction band very close to
\emph{half-filling} and a \emph{perfect lattice} of quantum impurities, coupled
through an antiferromagnetic Kondo exchange
$J_K$: This produces a Kondo lattice model (KLM).
As in higher-dimensions\cite{Aeppli}, this
results in a Kondo insulating phase\cite{Ueda_review}. 
Then, we may investigate
two different classes of
metal-Kondo insulating transitions occurring in 
low-dimensional KLM's. 

First, we study the commensurate-incommensurate 
transition arising in the {\it strong} Kondo coupling limit 
$J_K/t\gg 1$ --- $t$ is the hopping amplitude of electrons --- by 
analogy to Mott insulators. We show how the Kondo coupling plays the role of
a strong umklapp process for conduction electrons.

Second, we stay at half-filling and study the {\it weak} coupling limit
$J_K/t\ll 1$. 
Here,
metal-insulating transitions\cite{Giam1} arise rather
by the strong renormalization of the backward Kondo
exchange. The central idea of this part is therefore to show how
the interplay between electron-electron interactions in the LL 
and backward Kondo
scattering creates noteworthy phenomena in transport properties, in the
delocalized regime. 
For weak attractive interactions, as a precursor of 
superconductivity, 
this will produce both a large jump in the
Drude weight and a maximum in the d.c. conductivity, in the entrance of
the Kondo insulating regime. 
For repulsive ones, the LL yields prominent spin-fluctuations, that can
be easily pinned by local moments: Then the d.c. conductivity decreases
\emph{monotonically} with temperature, even at high temperatures. There is
no remnant of the original umklapp process --- driven by the
Hubbard term.

We also make comparisons
with the case of randomly distributed magnetic defects: This could
lead to predictions on persistent currents of mesoscopic rings with prominent
quantum defects.

The precise 
plan of the paper is organized as follows. In section II, we consider the
pure KLM in
the strong Kondo coupling limit, and show how a 1D Kondo insulator can be
understood as an umklapp becoming relevant. 
Implications on the resulting commensurate-incommensurate
transition are then considered. 
In section III, we present several Kondo insulators occurring in the
weak Kondo coupling limit, dependently on the interaction 
between local moments and their spins. We also draw the
phase diagram as a function of electron-electron
interactions. In the central section IV, we investigate transport properties
(a.c. and d.c. conductivities, Drude weight) in the high-temperature
delocalized regime, and make substantial links with non-magnetic 
and magnetic Gaussian-correlated disorders.
Finally, in section V, we link the two extreme cases, single- and many quantum
impurities.

\section{KLM in Strong coupling, and doping}

Let 
us start with the pure KLM in the strong Kondo coupling limit
$J_K\gg t$, at half-filling.
The weak original electron-electron interaction can be here
neglected. Moreover, in such limit, including a direct exchange J
between local moments does not
affect the results on charge properties at
the commensurate-incommensurate transition. So, we
ignore it as well and for simplicity \emph{in this section}, we consider 
local spins with S=1/2. For an explicit description
of such phase, consult ref.\cite{Ueda_review}.

One of the remarkable properties of the present spin-liquid phase is its
different energy scales in spin and charge degrees of freedom. 
Most clearly, it is characterized by a ratio
of the charge- and spin gaps, $\Delta_c/\Delta_S$ which is not equal to
unity. The charge gap is larger
by $50\%$ ($\Delta_S=J_K$ and $\Delta_c=3J_K/2$).

Of course, spin properties in
the 1D Kondo lattice model (KLM) are completely
different from those in the large $U$ Hubbard model. This is rather
described by a quasi
long-range Resonating Valence Bond 
wave function due to the absence of a spin gap in the spectrum. 
However, the 
origin of the charge gap is that when
another electron is added to a local singlet, it costs a large energy of
the order of $J_K$ by breaking the singlet. 
This process works as a strong effective
 on-site repulsion between original electrons of the order of
$J_K$. The mechanism 
for opening the charge gap
is completely identical to that in a Mott-Hubbard solid\cite{Schulz1,Giam1}.
Integrating out spin degrees of freedom (see Fig.1), a 
Kondo insulator occurs only as a result of 
commensurability: \emph{In 1D, this may manifest
in umklapp scattering becoming relevant}\cite{Schulz2}. In the following, we
explain precisely 
why a 1D Kondo insulator can be built identically to a Mott insulator.
\begin{figure}
\centerline{\epsfig{file=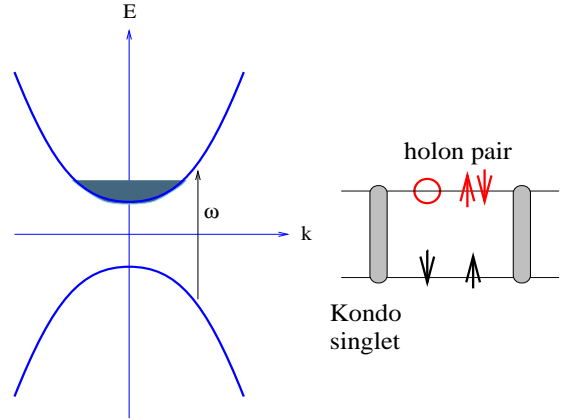,angle=0.0,height=5.5cm,width=7.3cm}}
\vskip 0.5cm
\caption{Picture of the KLM in strong coupling, with S=1/2. 
Separation of spin and charge
arises. When the hopping amplitude is very small,
spin excitations are optical magnons localized on the rungs. 
The lowest charge excitation consists of (breaking two singlets and) forming
a pair of (free) empty and doubly occupied sites with 
fermionic statistics, as in a Mott insulator at the Luther-Emery point.}
\end{figure}

\subsection{Mott insulator versus Kondo insulator}

In a 1D Mott insulator, as a result of
commensurability, the charge Hamiltonian is slightly more complicated
than the so-called Luttinger Hamiltonian 
\begin{equation}
\label{LL}
 {\cal H}_c=\frac{u}{2\pi}\int\ dx\ \frac{1}{K}:(\partial_x\Phi_c)^2:
+K:{(\Pi_c)}^2:,
\end{equation} 

and contains, in addition to
the quadratic part, a Sine-Gordon (SG) umklapp term
\begin{equation}
{\cal H}_{um}=\frac{2g_{3\perp}}{(2\pi a)^2}\int dx\ 
\cos\sqrt{8\pi}\Phi_c(x).
\end{equation}
The displacement field $\Phi_c$ and phase field field $\Theta_c$ satisfy
the usual commutation rules.
All interaction effects are taken into account through the velocity
$u$ and the Luttinger liquid parameter K (LLP). 
This obeys $K=1$ in the absence of interactions, $K>1$ for
attractive interactions and $K<1$ for repulsive ones. 
It is convenient to
perform the following canonical transformation:
$\Phi_c\rightarrow \sqrt{K}
\Phi_c$, $\Pi_c\rightarrow\Pi_c/\sqrt{K}$.
Then, ${\cal H}_{um}$ reads:
\begin{equation}
{\cal H}_{um}=\frac{2g_{3\perp}}{(2\pi a)^2}\int dx\ 
\cos\sqrt{8\pi K}\Phi_c(x).
\end{equation}
The interaction is then absorbed in the argument of the cosine term. For
repulsive interactions, this produces a quasiparticle gap $\Delta\propto
{g_{3\perp}}^{1/(2-\frac{n^2}{4\pi})}$, with $n^2=8\pi K$. 

\emph{A strong (on-site) repulsion between electrons must inevitably result
in $K=1/2$ close to half-filling}\cite{Schulz3,Anderson-Ren}. Then, we 
start with such bare value of the LLP.
Charge degrees of freedom (holons) should behave as
\emph{spinless fermions}\cite{remark0}. Through the Jordan-Wigner transformation in 1D (Appendix A), 
one can formally rewrite bosons operators in terms of spinless fermions 
$(q=\pm)$\cite{Haldane,Schulz1}:
\begin{equation}
\label{sept}
\psi_{q}(x)=\frac{\eta_q}{\sqrt{2\pi a}}:\exp
i \sqrt{\pi}(\Theta_c+q\Phi_c)(x):.
\end{equation}
This procedure is called refermionization.
Klein factors $\eta_q$ are built to fullfill anticommutation
rules between left $(-)$ and right $(+)$ movers.
More crucially, for $K=1/2$, the SG-umklapp definitely creates
fermionic kinks\cite{Schulz2}
\begin{eqnarray}
\label{mass}
\frac{g_{3\perp}}{(2\pi a)^2}\cos\sqrt{4\pi}\Phi_c &=& \frac{\Delta}{2\pi a}
\cos\sqrt{4\pi}\Phi_c \\ \nonumber
&=& -
i\Delta \hbox{\large(}
\psi_{+}^{\dag}\psi_{-}-\psi_{-}^{\dag}\psi_{+}\hbox{\large)}.
\end{eqnarray}
Klein factors have been chosen as $\eta_+\eta_-=+i$.
The mass term 
always favors the pinning of the $4k_F$ charge-density wave (CDW), producing in
our case a Kondo insulator, which is rather characterized by an even number
of electrons per unit cell when adding the local
moments. The holons 
disappear from the Kondo ground state. An excited holon from the valence 
band carries 
a wave vector $q=2k_F$. A pair holon-(anti)holon\cite{remark2} 
produces a kink of $2\pi$ in the
superfluid phase $\sqrt{\pi}\Theta_c$ at $x$, i.e. a kink of 2
in the charge current, that definitely
corresponds to a pair of empty and doubly occupied sites with fermionic
statistics. Fourier transforming, one finds:
\begin{equation}
\label{un}
{\cal H}_o =\sum_k (uk) qc^{\dag}_{q,k}c_{q,k}
+\frac{g_{3\perp}}{2\pi a}
\hbox{\large(}c^{\dag}_{+,k}c_{-,k}+h.c.\hbox{\large)}
\end{equation}
Using a so-called Bogoliubov transformation, this gives 
us the new energy spectrum:
\begin{equation}
E_{\pm}=\pm \sqrt{{u}^2k^2+{\Delta}^2}.
\end{equation}
At half-filling, the umklapp term gives us a semi-conductor picture of two
bands. Therefore, to recover the physics of the KLM in the limit 
$J_K/t\gg 1$, we have only to take the quasiparticle gap\cite{Ueda_review}:
\begin{equation} 
\Delta=g_{3\perp}/2\pi a=3 J_K/4. 
\end{equation}
Remarkably, charge properties of
half-filled KLM's remain qualitatively unchanged
tuning the Kondo interaction from the strong-coupling to the weak-coupling
limit (Section III and especially Appendix B).

 \emph{Here, 
at low temperatures, the Kondo interaction
can be exactly rewritten as an effective umklapp.}
However, $g_{3\perp}$ (typically
 the quasiparticle gap) has then a less explicit
dependence in $J_K$, due to the consequent renormalization of the 
backward Kondo coupling at half-filling. 

\subsection{Consequences of doping}

To study the influence of {\it hole} doping on transport properties, 
we now shift the chemical potential
at the top of the lower subband $(\mu_{ch}=-\Delta)$. We get:
\begin{eqnarray}
\label{zero}
{\cal H}_o&=&\sum_k u_c k(a^{\dag}_{+,k}a_{+,k}-a^{\dag}_{-,k}a_{-,k})\\ 
\nonumber
u_c&=&\frac{\partial E}{\partial k}=\frac{u^2
  k_c}{\sqrt{(uk_c)^2+\Delta^2}}\cdot 
\end{eqnarray} 
The fermions (holons) $a_{\pm,k}$ now refer to the partially occupied subband
and linearizing the band structure near the new Fermi points produces
$k_c=\pi\delta$ [$\delta=1-n$, being the hole doping]. 

By
doping such a semiconductor, 
low-energy properties should be well described through a Luttinger liquid
Hamiltonian with a velocity $u_c\sim \delta/\Delta$ and $K\rightarrow 1/2$.
In the 1D KLM below half-filling, such a LL behavior
has been checked numerically, for instance, in ref.\cite{Tsvelik_LL}.
This leads to important consequences. 

This produces a metallic phase with {\it finite} compressibility
$\kappa\propto K/u_c$ and Drude weight ${\cal D}\propto
u_c K$ at zero frequency. 
Very close to half-filling $(u_c\propto
\delta\rightarrow 0)$, then the Drude weight should
vanish continuously (with
a dynamical exponent $z=2$\cite{W-Kohn,remark1})
and $\kappa$ diverges as $\delta^{-1}$. 
A recent 
Density Matrix Renormalization-Group (DMRG) study has confirmed for large
$J_K$ parameters and $T=0$,
that $\kappa=\chi_c$ (charge susceptibility)
diverges as $1/\delta$ near half-filling\cite{Shibata}.
Moreover, at low temperatures, one then expects an exponential behavior
of the d.c. conductivity.
\vskip 0.1cm
\begin{figure}
\centerline{\epsfig{file=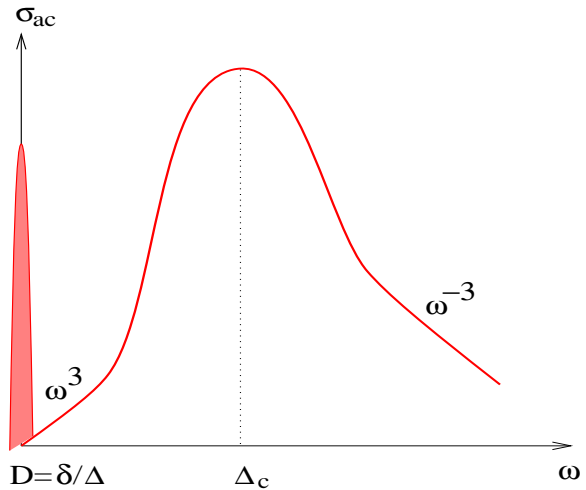,angle=0.0,height=6.4cm,width=7.6cm}}
\vskip 0.3cm
\caption{a.c. conductivity near the metal-insulator
transition induced by doping. In the strong coupling limit
$J_K/t\gg 1$, the spinless
fermion representation of holons or the umklapp-description 
should remain convenient even at quite high
frequency.}
\end{figure}  

The regular part of a.c. conductivity
must have two distinct regimes: an $w^3$ absorption at 
small frequency, and a tail $w^{-3}$ at large frequency\cite{Giam1,Giam3}.

To summarize, the commensurate-incommensurate transition
of the 1D KLM belongs to the
so called Prokovsky-Talapov class of universality\cite{Schulz2,pro}.
Now, let us push forward the duality with the Hubbard model in the
\emph{strong} $U$-limit, \emph{below} half-filling. 

The model
can be mapped onto a Luttinger Hamiltonian as long as $n\neq 1$.
Moreover, there is a remarkable duality by exchanging the
two parameters $n\leftrightarrow\delta$. Indeed,
the parameters $K$ and $u_c$ which determine the behavior
of the system are known to remain unchanged under the exchange 
$n\leftrightarrow \delta$\cite{Schulz1}. In particular 
$u_c=2t\sin(\pi n)=2t\sin(\pi\delta)$ is the {\it exact} velocity of the 
charge excitations for infinite repulsion\cite{Lieb-Wu}.
Actually, this implies 
that there are two equivalent holon representations in the
1D KLM, when $J_K\rightarrow +\infty$.
First, treating the introduced empty sites
as holons leads to a gas
of noninteracting spinless fermions in low density 
$\sim\delta$. As shown precedingly, this picture is particularly
accurate to capture transport properties at the commensurate-incommensurate
transition. Second, one could also identify the Kondo singlets as 
holons (single-occupied states would be then localized moments). 
This produces a gas
of noninteracting spinless fermions in large density $\sim n$\cite{Lacroix}. 
This picture has been preferred to show
the occurrence of {\it ferromagnetism}
between unscreened local moments, with an exchange coupling $\sim
-t^2/J_K$\cite{Sigrist,Troyer}. 

In the pure KLM, 
the unscreened localized spins polarize 
completely in order to
lower the kinetic energy of mobile holons. Such a case is a concrete
application of the Nagaoka's theorem in 1D. 
The physics behind this is the
finite extension of the screening cloud, which leads to the stability of the
ferromagnetic state at finite densities not restricted to the 
single-electron case. 

Note that including an antiferromagnetic Heisenberg 
exchange $J$ between local moments $J\gg t^2/J_K$
produces a so-called t-J model and
then rather a paramagnet\cite{Affleck}.

\subsection{Interacting Kinks}

Decreasing $J_K$ from infinity, results found precedingly on the
commensurate-incommensurate transition 
become now available only for 
\emph{very low} doping: Kinks become interacting below half-filling. 

Most clearly, a repulsive interaction
between the neighboring spinless fermions is introduced in the order
of ${\cal O}(t^2/J_K)$\cite{Ueda_review}. The situation becomes then similar
to the infinite $U$ Hubbard model with nearest neighbor
repulsion whose $K$ becomes {\it smaller} than 1/2\cite{Schulz3}.
More importantly, the ferromagnetic phase of the
pure KLM is very unstable decreasing $J_K$
due to the long-range overlap between Kondo singlets resulting in a certain
magnetic disorder (see next section). Precisely, 
a first order phase transition in the intermediate coupling
limit from the ferromagnetic state to a paramagnetic one
has been pointed out recently\cite{Tsvelik_LL}. The latter
is accompanied by a 
jump both in the magnetization curve and the value of 
$K$ (which curiously sharply decreases). This cannot be explained using
the preceding analysis. 

To conclude: the dependence of $K$ as a function of $J_K$
for weaker Kondo couplings remains difficult to handle \emph{below}
half-filling. A general trend, however,
is that $K$ is always smaller than 1/3 producing
a dominance of $4k_F$ charge oscillations, and then a so called
Wigner crystal. 

\section{Weak-coupling Kondo transitions}

Next, we subsequently stay at half-filling.
We explore the perturbative limit, $U/t\ll 1$ and
$J_K/t\ll 1$. Here, 
bosonization tricks allow us to explain the occurrence of Kondo 
insulators in commensurate systems when $J_K/t\ll 1$, in a simple way. 
\vskip 0.2cm
The small-$J_K$ regime
is of most interest, since correlation effects 
(for
instance correlations between local moments, or between electrons) 
are most important there.
For this regime, the expansion in $t/J_K$ is, of course, not convergent while
variational methods are not well controlled. This is mainly because 
$J_K=0$ is a singular limit leading to a metallic behavior. 
\vskip 0.2cm
To bosonize the Kondo interaction we need the bosonic representations for the
conduction electron- and the localized spin operators. \emph{Again, 
we limit the following study to the case of a single-channel conduction
band}. In a very general form, that can be used whatever the strength of
the Heisenberg interaction J between local moments or their spins S, these
are given by\cite{Affleck_boso}:
\begin{eqnarray}
\Psi^{\dag}_{\alpha}&(x)&\frac{{\sigma}_{\alpha\beta}}{2}\Psi_{\beta}(x)=
(\vec{J}_{R}+\vec{J}_{L})\\ \nonumber
&+&
\frac{e^{2ik_Fx}}{2\pi
a}\hbox{Tr}\{\vec{\sigma}(\Phi^{(1/2)}+\Phi^{(1/2)\dag})\}\cos{\sqrt{2\pi}
\Phi_c}\\ \nonumber
\vec{S}(x)&=&S\hbox{\large(}a\vec{L}(x)+(-1)^x
\sqrt{1-(a{L})^2}\vec{N}(x)\hbox{\large)},
\end{eqnarray}
where $|\vec{L}|a\ll 1$ is the quickly varying ferromagnetic component of the
local magnetization, and:
\begin{equation}
\vec{L}(x)\cdot\vec{N}(x)=0. 
\end{equation}
In the weak-coupling limit, fluctuations
in the $2k_F$ electronic spin operator are produced by the presence both, of
doubly occupied and empty sites [for small $U$, holons are
described in Appendix B] and of so-called
domain walls [spinons of the Heisenberg chain 
are introduced in Appendix C].
The Hamiltonian for electrons will be also decomposed into a holon
part [the same as Eq.(\ref{LL}) with $K=1-Ua/\pi v_F$] and the usual
spinon part. 
\vskip 0.15cm
Moreover, 
away from half-filling, the continuum limit
of the Hamiltonian only contains a marginal Kondo-coupling of the 
currents\cite{Affleck}:
\begin{equation}
\label{forward}
{\cal H}_{m}=\lambda_2[\vec{J}_{R}+\vec{J}_{L}]\vec{L}(x).
\end{equation}
At half-filling, the $2k_F$ oscillation becomes commensurate with the
alternating localized spin operator and the most prevalent contribution
reads:
\begin{equation}
\label{backward}
{\cal H}_{hf}\propto\lambda_3\cos{\sqrt{2\pi}
\Phi_c}
\hbox{Tr}\{\vec{\sigma}(\Phi^{(1/2)}+\Phi^{(1/2)\dag})\}\vec{N}(x),
\end{equation}
$(\lambda_2, \lambda_3)\propto J_K$ being the usual {\it forward} and 
{\it backward} Kondo scattering processes. Notice
 the close analogy with the two-chain spin system\cite{Dagotto-Rice}.

\subsection{Renormalization-group- and exact treatments}

First, it is convenient to use
the renormalization group method, expanding in powers of the coupling
constant $\lambda_3$. The perturbation is divergent, and one can derive
renormalization equations upon rescaling of the short distance cut-off 
$a\rightarrow a e^l$:
\begin{equation}
\label{reclaw}
\frac{d\lambda_3}{dl}=\hbox{\Large(}\frac{3}{2}-\frac{K}{2}-\Delta_N
\hbox{\Large)}\lambda_3,
\end{equation}
\emph{$\Delta_N\leq 1/2$, being in the following
the scaling dimension of the localized spin operator (see below, for specific
cases).} 

At half-filling, we expect
${\cal H}_{hf}$ to produce a gap for charge excitations [whatever
J or the spin S of local moments]. The charge gap should
follow the quasiparticle gap at the Fermi level
\begin{equation}
\label{gap}
\Delta\propto \lambda_3^{\frac{2}{3-K-2\Delta_N}},
\end{equation}
for either sign of $\lambda_3$. In the weak coupling limit,
a Kondo insulator is then formed 
due to the strong renormalization of the Kondo exchange $\lambda_3$. 

It should be noted that
the exchange coupling $\lambda_3$
and the usual umklapp $g_{3\perp}\propto U$ --- driven by the on-site
Hubbard term --- affect the Luttinger parameter $K$ in a symmetric
way:
\begin{equation}
\label{K}
\frac{dK}{dl}=-u K^2\hbox{\Large(}C_0g_{3\perp}^2+C_1\lambda_3^2\hbox{\Large)}
J_o(\delta(l)a).
\end{equation}
$J_o$ is the Bessel function, and $C_0$, $C_1$ are constants. 
At half-filling we must take $J_o(0)=1$.

When $\lambda_3=0$, the usual umklapp process 
is known to produce a Mott
transition at half-filling. It occurs for a critical value of K
which is equal to one, i.e. at the noninteracting point. 
As soon as $\lambda_3\neq 0$, the
physics becomes ruled by the Kondo coupling, producing a Kondo 
transition for a critical Kondo coupling $J_K^c=0$. 
The usual umklapp process 
can be neglected for small $U$ because it (only)
scales marginally to strong couplings.

Second, at low temperatures $T\ll\Delta$ the holons, becoming 
massive, also change
of statistics (semions $\rightarrow$ fermions). Charge carriers
behave as fermionic kinks due to the strong
renormalization of $J_K$, and as said before the backward Kondo
interaction can be exactly rewritten as an ``effective''
umklapp process (see Appendix B) --- remarkably, whatever $\Delta_N\leq 1/2$.
The charge gap will have, of
course, dramatic consequences for the physical properties. First, this implies
a long-range order in the $\Phi_c$-field. Indeed, we have
to take formally, $K^*=0$, for $T\ll\Delta$.
Then, at the Kondo transition, 
there is a finite jump in the compressibility and Drude weight. This will be
analyzed in more details in the next section.

\subsection{Spin properties for several $\Delta_N$'s}
 
Before investigating transport properties in great details, it is maybe
appropriate to review several interesting 
Kondo spin-liquid phases occurring in
the weak coupling limit, for particular values of
$\Delta_N$. 
Details of the technique can be found in Appendix C.
\subsubsection{$\Delta_N=1/2$: Heisenberg-Kondo lattice for S=1/2}
Let us start with the
so-called Heisenberg-Kondo lattice, where the spins 
of the array are coupled through a consequent
 antiferromagnetic coupling exchange J of the order of t. Here, one can
introduce spinon-pairs:
\begin{equation}
\vec{N}=\hbox{Tr}[({\Theta}^{(1/2)}+{{\Theta}^{(1/2)\dag}})\vec{\sigma}].
\end{equation}
For weak Hubbard interactions (for a parallel, see formula B3), one
gets also: $J_K<\cos\sqrt{2\pi}\Phi_c(x)>\ =\Delta$.
Therefore, the backward Kondo interaction is transformed
into:
\begin{equation}
\label{coset}
{\cal H}_{hf}=\Delta\hbox{Tr}{\bf{\Phi}^{(1)}}-3\Delta\epsilon.
\end{equation}
By analogy to usual spin ladder systems, the spin spectrum is composed of
massive triplet excitations [described
by $\hbox{Tr}{\bf{\Phi}^{(1)}}$] with a mass $m_t=\Delta$, and of a
high-energy singlet branch at $m_s=-3\Delta$\cite{Totsuka-Suzuki}.
We expect
the spin gap to decrease for any appreciable difference between J and t.

\subsubsection{$\Delta_N=0$: Weakly coupled S=1/2-local moments}

Precisely, let us now study the opposite limit
where the local moments are weakly coupled i.e. $J\ll J_K$.
At short distances, the RKKY interaction --- that displays a very small 
decay in
1D --- produces a perfectly static staggered potential with:
\begin{equation}
\vec{N}=\hbox{constant}.
\end{equation} 
The electrons are then subject to Bragg scattering\cite{KLH_Ising}. 
This opens a quasiparticle gap (linear in $J_K$). 
On the other hand, these screen away the internal field before a
true magnetic transition takes place. 

Then, there are still some {\it triplet} states in the 
quasiparticle gap because the SU(2)-spin symmetry is 
restored at long distances. The resulting spin gap becomes considerably
smaller than the charge gap\cite{Zachar_KLM}.

First, 
if J is not so far from $J_K$, these spin degrees of freedom are described in
terms of an O(3) non-linear sigma model without the topological term. The
implicit breaking of conformal invariance rescales the spin gap as:
\begin{equation}
\Delta_S\propto \Delta \exp-{\pi
  \hbox{\huge(}\frac{|J-t|}{t}\hbox{\huge)}}\cdot 
\end{equation}
Single electron excitations in 
the interval $\Delta_S$ and $\Delta$ can be viewed
precisely as polarons --- i.e. bound states of an electron with a 
\emph{kink} of the vector $\vec{N}$\cite{Tsvelik_charge}. 

Second, if J tends to 0, the condition for the 
presence of kink in the local staggered
magnetization operator now breaks down. Spin degrees of freedom rather behave
as follows\cite{KLH_Ising}. At not too long distances,  
half of the (electronic) spinon field begins
to fluctuate due to the restoration of the
SU(2)-symmetry, contributing to a new chiral phase of the same
universality class as the 2D Ising model. 
At long distances, due to
strong fluctuations in the spin array, the (electronic)
spinon field cannot be pinned anymore by
the backward Kondo coupling --- i.e. $|\vec{N}|=0$.
Then, spin flips should contribute. The system may scale to a 
perfect singlet ground state at a very low energy scale:
\begin{equation}
\label{impurity}
\Delta_S\propto\Delta\exp-2\pi v_F/\lambda_2\ll\Delta,
\end{equation} 
typically the single-site Kondo temperature, in
agreement with numerical calculations\cite{Ueda_review}.

To conclude this part: For spin-1/2 local moments, 
the ground state is always a spin
{\it singlet}. Furthermore, for small
$J_K$'s, the antiferromagnetic correlation length is quite large producing
disordered quantum fluids. It is already increased
by an appreciable difference $(J-t)$, and becomes huge for J=0.

\subsubsection{$\Delta_N=3/8$: Underscreened S=1-chain}

Now, we consider another generic
case where the local spins form a so-called
S=1 Takhtajan-Babujan chain. Details of the calculations can be found 
in ref.\cite{KLH_underscreen}. These can be extended to the case of a spin 
S-integrable chain $(|1-2S|\neq 0)$ --- with
$C_v/L=2ST/(1+S)$\cite{KLH_unpublished,NE}.

In the sense of critical theories, this model
can be parametrized by a conformal anomaly $C=3/2$. In the
continuum limit, this spin-1 chain has then only zero sound {\it triplet}
excitations --- described by three massless Majorana fermions ---
or rather magnons. The staggered magnetization 
$\vec{N}$ has now dimension $\Delta_N=3/8$. Consult Appendix C3.
The Kondo interaction grows to strong
coupling, producing a quasiparticle gap $\Delta\propto
{J_K}^{8/5}$. 

Then, a new phenomenon arises: The occurrence 
of singlet bound states can be
achieved only due to the ``fractionalization''
 of each spin triplet excitation onto 
{\it two}
spins-1/2. On each rung, (only)
one is strongly coupled to the conduction band. This
produces the same contribution as in Eq.(\ref{coset}), and then \emph{optical
magnons} in the system. From a general point of view, Luttinger
liquids coupled to an active insulating
environment via a Kondo-like coupling yield
a spin gap\cite{Granath-Johannesson}. 
But, {\it deconfined spinon-pairs} still
subsist in the ground state because
the local moments are underscreened. 
 This leads to an anomalous optical conductivity at low frequencies, and then a pseudogap
phase. Starting with small J's (the Haldane
gap is irrelevant), predictions from the non-linear sigma model for S=1
give the
same conclusion\cite{Tsvelik_charge}.

\subsection{Role of electron-electron interactions}

The present study shows that the smallest amount of Kondo potential
$\lambda_3$ produces a Kondo insulator at least for sufficiently large 
length scales. Then, the backward Kondo scattering can be naturally
rewritten as an effective \emph{strong} 
umklapp process, such that the original umklapp
term --- again, driven by the Hubbard interaction --- can be neglected.

One can check that the charge gap increases with (enhancing) $U$ --- or
decreasing the LLP. If the Hubbard repulsion is very large $U/t\gg 1$ 
one can expect to see a crossover from a Kondo insulator to a usual
spin ladder system\cite{Dagotto-Rice}. Here, the insulating transition should
be rather driven by U only, and then we have to take formally
$<\cos\sqrt{2\pi}\Phi_c(x)>\ =\sqrt{U}=\hbox{constant}$
and $K=0$ in the spin gap equation.
For instance, for $\Delta_N=1/2$, we recover previous results and for
instance the spin gap $\Delta_s\propto J_K$ (see Appendix C).
\vskip 0.05cm
\emph{Another interesting feature is that the
Kondo insulating state should even persist in presence of attractive 
interactions}. 

A physical mechanism for the generation of attractive
interactions is electron-phonon interaction. A renormalization group
treatment of the important spin-backscattering term $g_{1\perp}$ 
(given in Appendix D) predicts, for $J_K=0$, the existence of a
spin gap
\begin{equation}
\Delta_{ss}\sim \frac{v_F}{a}\exp\hbox{\Huge(}\frac{-\pi
  v_F}{\left|U\right|a}\hbox{\Huge)}\cdot
\end{equation}
Remarkably, localization by magnetic impurities remain the most prominent 
phenomenon
as long as $K<K_c=3-2\Delta_N$. When $K=K_c$, the Kondo gap tends to zero
and then, a superconducting-like ground state takes place.
When $K\geq K_c$, correlation functions
for the singlet superconducting pairing field
are considerably increased in the
low-temperature phase $(T\ll\Delta_{ss})$ and the charge sector remains
critical, leading to a perfect superfluid. 
This produces a coexistence between a Luther-Emery liquid and
an insulating part characterized by 
strong antiferromagnetic fluctuations.
The pairing also promotes $2k_F$ CDW
fluctuations that are not sensitive
to magnetic impurities.
\vskip 0.2cm
\begin{figure}
\centerline{\epsfig{file=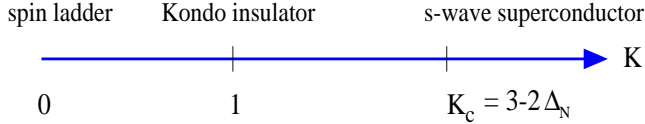,angle=0.0,height=1.65cm,width=8.5cm}}
\vskip 0.3cm
\caption{The phase diagram as a function of the LLP, $K$. For a large range of
$K$ parameters (corresponding to weak $U$'s), 
the interaction between electrons
and magnetic impurities is well prominent, producing a
Kondo insulator. For attractive enough interactions $(K>K_c)$
the electron liquid can turn into a s-wave superconductor 
decoupling completely from the spin array.
A large coherence between spins of the array, resulting
in $\Delta_N=0,$ tends to increase $K_c$.}
\end{figure}
\vskip 0.1cm
To conclude: Close to the superconducting
transition $(K<K_c)$, as a remnant of
s-wave superconductivity (and CDW fluctuations)
one expects
that the electron liquid is {\it less} localized 
than for repulsive interactions.
Let us now investigate transport properties and comment this point
in more details.

\section{Transport properties}

Transport in such commensurate systems is very interesting because the
Kondo process $\lambda_3$ provides an important relaxation mechanism for
electrons. Next, we will assume that the coupling $\lambda_3$
is sufficiently \emph{weak}
 that some perturbative calculation of the conductivity
as a function of $\lambda_3$ can be performed. The conductivity itself
does not have a regular expansion in $\lambda_3$. A way out is provided by
the memory function formalism. 

If the system is a normal metal
with finite dc-conductivity, one can define\cite{Gotze-Wolfle}
\begin{equation}
\label{sigma}
\sigma(w)=\frac{2iK u}{\pi}\frac{1}{w+M(w)}\cdot
\end{equation}
and $M(w)$ is the meromorphic memory function. This formula
is well-suited for an ``infinite'' system: \emph{Next, 
we are not interested in reservoir effects}.

\subsection{Memory function approximation}

The calculation of
this function can be carried out perturbatively to give at the lowest order
\begin{equation}
\label{mf}
M(w)=\frac{(\ll F;F\gg_w-\ll F;F\gg_{w=0})/w}{-\chi(0)}\cdot
\end{equation}
At zero frequency, one gets the retarded current-current correlation function: 
\begin{equation}
\chi(0)=-2u K/\pi. 
\end{equation}
The factor 2 comes from the two colors of spin which both contribute
to transport properties.
The symbol $\ll;\gg$
designs a retarded correlator computed in the absence of magnetic 
impurities\footnote{As long as $\lambda_3$ remains small, one can neglect all effects of
self-adjustments of the ground state to the Kondo potential.}
and $F=[J_c,H]$, $J_c$ is the charge
 current. The F operators take into account the
fact that the charge
current is not a conserved quantity. 

For frequencies $w\gg M(w)$ and $T\rightarrow 0$, one gets:
\begin{equation}
\label{sigma-ac}
\sigma_{ac}(w)=\frac{-i\chi(0)}{w}[1-\frac{M(w)}{w}]\cdot
\end{equation}
The regular part in the delocalized phase reads:
\begin{equation}
\sigma_{reg}(w)\propto M(w)w^{-2}.
\end{equation}

The expression (\ref{mf}) does not necessarily remain valid at very
low frequencies,
even for finite temperatures. Its validity at low frequencies implicitly
assumes in a self-consistent way that the d.c. conductivity behaves as:
\begin{equation}
\sigma_{dc}(T)=\frac{-i\chi(0)}{M(T)},
\end{equation}
and $M(T)$ is related to some relaxation time, i.e.
\begin{equation}
\label{memory}
M(T)\propto \tau_{rel}^{-1}. 
\end{equation}
In the macroscopic pure system, one gets $\sigma_{dc}(T)\rightarrow +\infty$.
{\it Here, $\tau_{rel}$ should correspond to the effective 
elastic time between two magnetic diffusions.} Of course, it will be highly
non-universal and temperature-dependent (since 
we are in a ballistic transport). This should legitimize the memory 
function approximation, which gives
in general good results as long as we stay far away from an insulating
region.
This approach, of course, will break down when $T=\Delta$ i.e. when the
effect of $\lambda_3$ on the ground state becomes strong: it opens a 
spin- and charge gap changing drastically the nature of the elementary
excitations. At low temperatures, 
the source of large resistance in such magnetic systems comes
from prominent quantum scattering which merely
destroys the coherence or the propagation 
of the accelerated preexisting excitations, producing Kondo localization.

Interestingly, here the quantum 
correlation between spins of the array --- driven by $\Delta_N$,
that plays a crucial role in the classification
of the Kondo localizations (e.g. gap
equation, spin spectrum), will explicitly appear in the computation of the 
high-temperature d.c. conductivity.
\vskip 0.1cm
The bosonization scheme gives:
\begin{equation}
\label{current}
J_c\sim\sqrt{2}uK\int \Pi_c(x)\ dx.
\end{equation}
Using the definition of ${\cal H}_{hf}$, the behavior
of the memory function as a function of temperature can be
easily obtained. We find:
\begin{equation}
\ll T_{\tau} F(\tau)F(0)\gg\ \propto \int dx\ {\lambda_3}^2
{\hbox{\Large(}\frac{1}{x^2+(u\tau)^2}\hbox{\Large)}}
^{\sqrt{1+K+2\Delta_N}}\cdot
\end{equation}
The function in the integral describes the amplitude that a
spinon and a holon recombine to give an electron
at the point $x$ and time $\tau$, that is immediately
diffused by an impurity.
Fourier transforming, one finds: 
\begin{equation}
M(w)\propto w^{(K-2+2\Delta_N)}, 
\end{equation}
and the same behavior as a function of temperature.
In the delocalized regime
$T\gg\Delta$ or $w\gg\Delta_c=2\Delta$, the temperature and frequency dependent
conductivity is
\begin{equation}
\label{cond}
\sigma_{dc}(T)\propto T^{2-\mu},\sigma_{ac}(w)\propto
w^{\mu-4},\mu=K(T)+2\Delta_N.
\end{equation}
For more details concerning the method, consult ref.\cite{Giam3}.

Away from the transition, one may consider the parameter
$K(T)$ as a constant $\sim K$.
We stress on the fact that dressing of the Kondo exchange by the other
interactions results in a nonuniversal power-law dependence in the delocalized
regime. Again, the whole perturbative scheme breaks down when 
$\lambda_3\sim 1$, at a length
scale which corresponds to the localization length of the system. 
A reasonable
guess of the temperature dependence below this length scale is an
exponentially activated conductivity because current-current correlation
functions now decrease exponentially in time\cite{Andrei-Rosch}. 

As can been seen from Fig.4, there is a
 maximum in $\sigma_{dc}$ as a function of T
for weak attractive interactions $(2-2\Delta_N<K<K_c=3-2\Delta_N)$, occurring
when the thermal coherence length is of the order in magnitude of the
localization length $T\sim\Delta$. 
{\it This maximum can be well understood as a
 remnant of s-wave superconductivity} in the localized phase
that cannot be easily pinned
 by magnetic impurities. Remind that the system is a
very good conductor until one reaches the localization 
temperature $\sim\Delta$.

A similar
phenomenon has been predicted in a disordered two-leg Hubbard ladder 
system with
attractive interactions\cite{Orignac}. In that case, `$2k_F$' 
charge density fluctuations are forbidden due to the spin gap, and
`$4k_F$' charge density
fluctuations, that aim to arise for repulsive interactions, become
very small for attractive ones\cite{Schulz}: this is
an important consequence of the
s-wave scenario in two-leg ladder systems. The nonmagnetic disorder can
become efficient only at very low energy.
For a comparison with the single Hubbard chain problem, see next subsection.
\vskip 0.3cm
\begin{figure}
\centerline{\epsfig{file=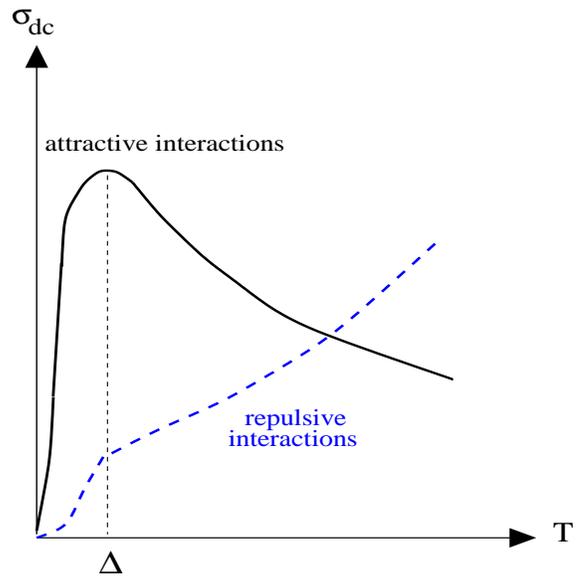,angle=0.0,height=7.5cm,width=7.5cm}}
\vskip 0.5cm
\caption{The behavior of d.c. conductivity as a function of temperature,
for repulsive and for attractive interactions. For $T\gg\Delta$,
$\sigma\propto
T^{2-K-2\Delta_N}$: For repulsive (electron-electron)
interactions, there is no maximum in
the conductivity and therefore no rest of original umklapps driven by the
Hubbard interaction. When the interactions
become attractive, the spin density fluctuations are suddenly lowered
making the system less localized by magnetic impurities than for
the repulsive
counterpart.}
\end{figure}
\vskip 0.03cm
For repulsive interactions, we can check that
there is no remnant of the original
umklapp term --- again, driven by the Hubbard
interaction ---
which is known to produce $\sigma_{dc}\propto T^{-1}$ (with $K\sim 1$) 
in the delocalized
regime, and then a maximum in $\sigma_{dc}$ in the entrance of the localized
phase. In the case of electron-electron interactions, the
relaxation time should result from
the coupling of electrons with a
thermal bath which, strictly speaking, is needed for the system to reach
thermal equilibrium\cite{A-R}. In such case, the relaxation time is 
rather associated to phase-breaking
or {\it inelastic} processes\cite{Giam3}. In particular, note that 
$\sigma_{dc}\equiv\tau_{in}$ (see Appendix D).
Furthermore, as long as a spinon is not scattered (as it is the case in the
absence of impurities), 
it may combine locally with another
holon to form an electron and the material is still a very good conductor. 
This is a consequence of spin-charge separation\cite{Anderson-Ren}.

Most clearly, for repulsive interactions, the ground state of the Hubbard chain
is almost a spin-density wave
(SDW) (with a power-law decay of the correlation functions) whose
charge density is uniform. 
{\it Such a ground state couples very strongly to magnetic
impurities} leading to prominent elastic processes in the
electron liquid. It is therefore not surprising that for
repulsive interactions the main source of resistance for the
electron liquid is due to the prominent scattering of 
a pair spinon-holon (electron) by a magnetic impurity. 

Furthermore, 
the conductivity is found to
decrease monotonically with temperature, even at high temperature.
This noteworthy renormalization of exponents and the faster decay of the
conductivity for repulsive interactions 
is a perfect signature of Kondo localization. 
Again, note the nonuniversality of exponents due to the factor
$\Delta_N$ which depends on the spin S of local moments and
 on the strength of the
Heisenberg interaction between them.
\vskip 0.2cm
To summarize the main point: 
Now, we are really able to allege that for repulsive interactions
the electron liquid
is more localized by the magnetic impurities
than for attractive interactions, both
because of the scale of localization and because of the d.c. conductivity
temperature dependence.

\subsection{Duality to an effective Gaussian disorder}

It should be noted
that these results can be reproduced defining the (Kubo) 
d.c. conductivity as: 
\begin{equation}
\sigma_{dc}(T)\sim T^{-1}/D_m(T)\hskip 0.3cm \hbox{and}\hskip 0.3cm
 D_m(T)={\lambda_3(T)}^2.
\end{equation}
$n(T)=1/T$ denotes the electronic density at a temperature T. We must
stop renormalization procedure at the thermal length $L=l_{in}=v_F/T$ at which
inelastic and decoherent effects take place.
By comparison with the case of a nonmagnetic Gaussian disorder\cite{Helene}, 
\begin{equation}
D_m(T)=l_{in}/l_e\propto T^{-(3-\mu)}.
\end{equation}
The exponent $\mu$ has been defined in Eq.(\ref{cond}).
In KLM's, $D_m$ will be identified as the {\it magnetic disorder} parameter. 
For more details, consult Appendix D. This definition traduces 
that at high temperatures we have a perfect duality to a
model of a 1D electron liquid submitted to a Gaussian correlated
spin disorder, with an effective
exponent $\mu=K(T)+2\Delta_N$ for the electronic
spin density-spin density correlation function. 
In the case of a Gaussian correlated {\it nonmagnetic} disorder, for SU(2)
symmetry and {\it repulsive} interactions one rather
expects\cite{Giam2,Peschel}:
\begin{equation}
\label{dis}
\sigma_{dc}(T)\propto T^{2-\hat{\mu}}\hskip 0.3cm \hbox{with}\hskip 0.3cm 
\hat{\mu}=1+K(T).
\end{equation}
Far away from the localization energy scale
i.e. for very small disorder, one can take $\hat{\mu}=1+K$\cite{Apel}.
\vskip 0.15cm
Note that to couple to nonmagnetic disorder
one has to distort the SDW and make a fluctuation of the charge
density, a process
which costs an energy of order U. 
Therefore, for repulsive interactions, 
the non-magnetic disorder effect must be 
inevitably weaker than for attractive interactions where
$2k_F$ CDW fluctuations can get very easily pinned
by nonmagnetic impurities. Starting with the Hubbard model
and \emph{small U}, the 
d.c. conductivity found above is then not completely correct. The
reason is that the backscattering spin
term $g_{1\perp}$ couples to the non-magnetic disorder, and
then it cannot be ignored (see Appendix D). At intermediate length scales, 
this results rather in:
\begin{equation}
\label{conductivity_nm}
\sigma_{dc}(T)\propto T^{-Ua/\pi v_F}.
\end{equation}
Now, the maximum in the d.c. conductivity
in the entrance of the Anderson glass phase occurs 
for repulsive interactions.
{\it The situation becomes opposite to that found 
with an array of magnetic impurities.}

In general, for repulsive interactions one gets 
$\mu<\hat{\mu}$, revealing that 
Kondo localizations may still take place in {\it weakly} disordered 
1D Kondo lattice models. Especially when $\Delta_N=0$, we
have a constant {\it uniaxial} Kondo potential at high temperatures
\begin{equation}
\overline{V(x)}=\lambda_3,\qquad \overline{V(x)V(0)}=D_m,
\end{equation}
and $\mu=K(T)\ll\hat{\mu}$. There is no
quantum fluctuation.
The long-range coherence of the spin array
produces inevitably Kondo localization.
On the other hand, adding explicitly a 
dominant antiferromagnetic direct exchange between local moments
should produce a stronger
competition between Anderson- and Kondo localizations, even if
the ground state is always of Kondo type for prominent 
repulsive interactions\cite{Karyn1}. 
This comes from the facts that 
$2k_F$ CDW fluctuations are not
completely forbidden at high temperatures
for {\it weak} on-site repulsive interactions, and 
that the Kondo potential follows:
\begin{equation}
\overline{V(x)}=0,\qquad \overline{V(x)V(0)}=D_m/x^{2\Delta_N}.
\end{equation}
It is now short-range correlated, and we have
$\Delta_N=3/8$ (Takhtajan-Babujan S=1 chain), or $\Delta_N=1/2$ (Heisenberg
S=1/2 chain). The average is explicitly
performed on quantum fluctuations (after normal ordering).
\vskip 0.1cm
Below, we will 
investigate the case of a Gaussian
correlated magnetic disorder (subsection E): 
again it will lead to the same conclusion.
Using Appendix D, one can easily
check that repulsive interactions tend to make the nonmagnetic disorder less
relevant and decrease localization. This has dramatic consequences
on the effects of interactions on persistent currents\cite{Giam-Shastry}. 
Comparisons with the renormalization of charge stiffness in 1D
KLM's  will be made later in subsection D.

\subsection{Charge incompressibility with $\sigma(w,k=\pi)\neq 0$}

In the following, we present a new class of Kondo insulators 
which
is (charge) incompressible, but yields only a pseudo gap in the 
(charge) optical conductivity at a large wave-vector $k=\pi$.

On the one hand, the Kondo transition at 
half-filling produces an incompressible system because $\kappa\propto K^*=0$.
On the other hand, ordinary Kondo insulators [where
local moments are characterized by $S=1/2$] have
no charge conductivity for low frequencies. First, 
the charge current $J_c$ given by the expression
(\ref{current}) vanishes. 
Second, the presence of a spin gap in the system produces
optical magnons, that doesnot allow any anomalous charge current at
large wave-vectors (see Appendix C). 

Now, let us investigate the case where local moments form a 
Takhtajan-Babujan S=1 chain. As shown precedingly,
free spinons remain in the ground state of the system due
to the `underscreening' of the local impurities. This implies that
the optical {\it charge}-conductivity would reveal {\it no gap} 
at low frequency and $k=\pi$. 
\vskip -0.05cm
\begin{figure}
\centerline{\epsfig{file=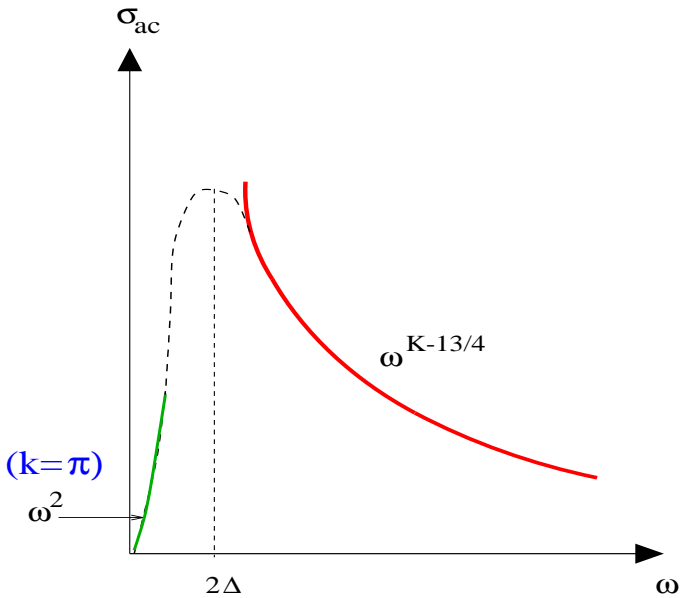,angle=0.0,height=7.8cm,width=9cm}}
\vskip 0.5cm
\caption{The behavior of a.c. conductivity at zero temperature as a function
of frequency in the case of a Takhtajan-Babujan chain
i.e. $\Delta_N=3/8$. Conversely to usual Kondo insulators,
$\sigma_{ac}(w,k=\pi)\propto {w}^2$ for vanishing frequencies. This is a nice
example of pseudo-gap phase.}
\end{figure}

Spinons
are governed by an extra term in the action [we use the duality: $\Phi^{(1/2)}=i\vec{\sigma}{\bf \Phi}$],
\begin{equation}
\Gamma(i\vec{\sigma}{\mathbf\Phi})=
\frac{i}{8\pi}\int dxd\tau\ \epsilon_{\mu\nu}\hbox{\large(}{\mathbf\Phi}
\cdot[
\partial_{\mu}{\mathbf\Phi}\times\partial_{\nu}{\mathbf\Phi}]\hbox{\large)}=
i\pi Q
\end{equation}
where $Q=1=\int dx\ q(x)$ is the associated topological charge. From
this (topological) charge, one
can define an anomalous current density\cite{Tsvelik_charge},
\begin{equation}
\bar{\psi}(i\sigma^z)\psi={\cal J}_S=
\epsilon_{\mu\nu}\hbox{\large(}{\mathbf\Phi}
\cdot[
\partial_{\mu}{\mathbf\Phi}\times\partial_{\nu}{\mathbf\Phi}]
\hbox{\large)},
\end{equation}
which has the scaling dimension 5/2. In this way, the charge- and the
spin sector are related to each other. Then, the pair correlation function
of $J_S$ induces an unusual feature in the optical conductivity at 
a large wave-vector $k=\pi$. {\it The anomalous current is nothing but the ordinary
current at $k=\pi$.}
We find a retarded current-current
correlation function $\chi(w)\sim -iw^{3}$ and then an optical conductivity,
\begin{equation}
\sigma(w,k=\pi)=\frac{i}{w}\chi(w)\propto w^2.
\end{equation}
We obtain a new class of
incompressible Kondo systems, with no gap in the optical conductivity
(Fig.5). Similar conclusions can be 
reached in the case where the direct
exchange between local moments is very weak\cite{Tsvelik_charge}. 

It should be noticed 
that a new phase of disordered commensurate insulators 
yielding ``similar'' features, has been pointed out 
recently\cite{Edmond-Thierry-Pierre}. The corresponding phase 
with $\kappa=0$ and $\sigma(w,k=0)\propto w^2$
results from
a strong competition between Mott- and Anderson localization
which leads to unusual {\it excitonic} effects. This requires
some interactions of finite extent.

\subsection{Renormalization of the Drude peak}

In addition to the regular frequency
dependence of the a.c. conductivity found precedingly, one can
compute the charge stiffness ${\cal D}$ of the macroscopic
system, which measures the
strength of the Drude peak (occurring at zero frequency):
\begin{equation}
\sigma_{ac}(w)={\cal D}\delta(w)+\sigma_{reg}(w).
\end{equation}
This comes from the first term in Eq.(\ref{sigma-ac}).
For fermions with spins, the conductivity
stiffness at length $l$ can be obtained using:
\begin{equation}
{\cal D}(l)=2u(l)K(l),
\end{equation}
with ${\cal D}(l=0)=2uK=2v_F$.
Again the factor of two comes from the fact that there are twice degrees
of freedom compared to the spinless fermionic case. The velocity $u$ obeys
a recursion law of the type:
\begin{equation}
\label{u}
\frac{du}{dl}=-C_2u^2K D_m(l).
\end{equation}
$C_2$ is a constant. From Eqs.(\ref{K}),(\ref{u}) 
one can easily obtain the renormalization
group equation for ${\cal D}$:
\begin{equation}
\frac{d{\cal D}}{dl}=-CD_m(l),
\end{equation}
and C is a constant. Using the flow of $D_m(l)$ given explicitly in Appendix
D, one obtains:
\begin{equation}
\frac{d{\cal D}}{dl}=-CD_m(0)e^{(3-\mu)l}.
\end{equation}
By stopping the renormalization procedure at a length
$L=e^l=v_F/T$, it follows:
\begin{equation}
\label{jumpa}
{\cal D}(L)-{\cal D}(1)=-C\frac{D_m(1)}{(3-\mu)}\hbox{\Large[}
L^{(3-\mu)}-1\hbox{\Large]}.
\end{equation}
We can notice that for repulsive interactions the Drude peak 
{\it fastly} decreases with $L$, that is also typical of a
Kondo localization.
For attractive interactions, the Drude
peak decreases more slowly and the renormalization of ${\cal D}$ stops
completely as soon as $\mu=3$ that coincides
with $K=K_c$. This is in agreement
with the fact that for $K\geq K_c$
 the system behaves as a perfect superfluid 
at low energy.
\vskip 0.2cm
In the localized phase $(T\ll\Delta)$, the opening of the gap produces
formally $K^*=0$ and then the Drude weight
vanishes. Using Eq.(\ref{jumpa}) for $L\rightarrow{\Delta}^{-1}$, 
one can therefore deduce that there is (also) a jump
in the Drude weight at the transition. Using
ref.\cite{remark1} and replacing $\delta$
by $(T-\Delta)$, one then finds that the dynamical exponent is $z=1$,
that is specific of a Kosterlitz-Thouless transition.
\vskip 0.2cm
Again note the close parallel with the calculation of the charge stiffness
in presence of a Gaussian disorder\cite{Giam-Shastry}:
\begin{equation}
{\cal D}(L)-{\cal D}(1)=-\hat{C}\frac{D(1)}{(3-\hat{\mu})}\hbox{\Large[}
L^{(3-\hat{\mu})}-1\hbox{\Large]}.
\end{equation}
Nonetheless, physical consequences are very different. For small U one gets:
\begin{equation}
(3-\hat{\mu})=1-\frac{Ua}{\pi v_F}\cdot
\end{equation}
For finite non-magnetic disorder, one
finds that the charge stiffness of the system for
repulsive interactions is larger than the one of the system for attractive
interactions. Again, SDW fluctuations that are prominent for repulsive
interactions tend to make the quenched
disorder less relevant and decrease localization. The charge density tends to
be rather homogeneous.
Conversely,
for attractive interactions, $2k_F$ 
CDW {\it fluctuations} can get very easily immobilized
by non-magnetic impurities.

\subsection{Randomly distributed magnetic defects}

For completeness, we now make links with the case where magnetic impurities
would be randomly distributed. To simplify at
maximum the discussion, we consider here
that these exercise 
a {\it unixial} random magnetic field along the
z-axis. One gets the Hamiltonian (see Appendix D):
\begin{equation}
H_{m.dis}=\int dx\ V(x)\cdot\rho_s^{z}(x),
\end{equation}
$V(x)$ (complex) is assumed to be Gaussian correlated, resulting in
\begin{equation}
\overline{V(x)}=0,\qquad \overline{V^*(x)V(0)}=D_m\delta(x).
\end{equation}
The forward scattering potential, that does not affect transport properties,
will be omitted. Now, we apply standard renormalization-group methods for
one-dimensional disordered systems\cite{Giam2}. In the 
Abelian representation, the random Kondo potential reads:
\begin{equation}
H_{m.dis}=\int dx\ V(x)\sin{\sqrt{2\pi}\Phi_s(x)}\cos{\sqrt{2\pi}\Phi_c(x)}.
\end{equation}
Here, random
magnetic impurities are not {\it explicitly} linked to
each other (the Kondo potential
is `on-site' correlated). 
Now, we show that preceding conclusions
remain unchanged: The electron liquid is more localized by magnetic impurities
for repulsive interactions than for attractive ones. We follow exactly the
same scheme as in Appendix A of ref.\cite{Giam2}. 
From a perturbative expansion in the magnetic disorder 
parameter $D_m$ and the spin backscattering $g_{1\perp}$, the result is:
\begin{equation}
\frac{dD_m}{d\ln T}=-(3-K_s-K+\frac{g_{1\perp}}{\pi v_F})D_m.
\end{equation}
The main difference with non-magnetic disorder comes from the sign (plus
and not minus) of the last term, coupling interactions and Kondo
potential. This is
because electronic spin degrees of freedom are linked to the random
field via a {\it sinus} term (and not a cosine). For a simple sight, consult
Appendix E.
As a consequence, 
we still have a strong competition between localization of the electron
liquid by (disordered) magnetic potential and superconductivity: For
$g_{1\perp}\rightarrow -\infty$, the flow of $D_m$ scales to zero. 
For small U, one finds:
\begin{equation}
\frac{dD_m}{d\ln L}=(1+\frac{Ua}{\pi v_F})D_m.
\end{equation}
and,
\begin{eqnarray}
\frac{dK}{dl}&=&-C_1 uK^2D_m,\\ \nonumber
\frac{dK_s}{dl}&=&-C_3{K_s}^2 D_m.
\end{eqnarray}
$C_3$ is a constant. In the infra-red region, quantum
coherence between impurities definitely appears and then
$K^*$ and $K_s^*$ approach {\it zero}. The spin backscattering term
$g_{1\perp}$ remains small. Furthermore, 
the $2k_F$ electronic spin density wave will be {\it weakly} pinned
by the random magnetic field --- induced by local moments --- resulting in a 
glassy- and presumably a disordered Ising phase\cite{Doty-Fisher}. 

Far in the delocalized regime, one can equate the exponent $\mu$ to 
$1+K$. The result is then:
\begin{equation}
\sigma_{dc}\propto T^{Ua/\pi v_F}.
\end{equation}
At short length scales, the impurities can be
considered as \emph{independent} (see next section).
One can also notice
the close correspondence with the so-called Heisenberg-Kondo
lattice model with $\Delta_N=1/2$. This fact certainly
traduces that in the delocalized regime
there is no fundamental difference between
on-site- and (very) short-range correlated Kondo potentials.
Likewise, a maximum in the d.c. conductivity curve again should take place in 
the
entrance of the Kondo (glass) phase for attractive
interactions. For an explicit comparison with non-magnetic impurities, consult
Eq.(\ref{conductivity_nm}). Finally, one can also check that repulsive
interactions help in reducing the conductivity stiffness, while attractive
interactions reduce the decrease of conductivity stiffness by magnetic
disorder
\begin{equation}
{\cal D}(L)-{\cal D}(1)=-C\frac{D_m(1)}{(3-\mu)}\hbox{\Large[}
L^{(1+\frac{Ua}{\pi v_F})}-1\hbox{\Large]}.
\end{equation}

On the other hand, the stiffness ${\cal D}$ generally
provides a nice measure of the persistent current in a mesoscopic ring 
induced by a small magnetic flux\cite{Buttiker}.
Therefore, the above results could have some applications in nearly
one-dimensional mesoscopic rings (of size L: The electron has to stay coherent
along the whole ring)
with \emph{prominent} magnetic defects: for
example, repulsive interactions should {\it decrease} persistent currents.
Extensions of such model will be considered elsewhere.

\section{Links with the single impurity case}

Let us now compare with the case of a single magnetic impurity in
a Luttinger liquid. At half-filling, here
usual umklapps produce an Heisenberg chain. The
resulting Kondo effect has been studied in ref.\cite{Eggert}. Away from
half-filling, the Kondo
interaction gives the same two 
contributions as in Eqs.(\ref{forward}),(\ref{backward}) [forward
and backward scatterings].
Since scattering occurs only on a
finite part of the sample, the conductance is the most appropriate way to
describe transport. 

\subsection{New renormalization flow}

For weak $J_K$'s, one can use the same renormalization method
expanding in the interactions $\lambda_2$ and $\lambda_3$. One obtains the
RG equations
\begin{eqnarray}
\label{ali}
\frac{dK}{dl}&=& 0\\ \nonumber
\frac{d\lambda_2}{dl}&=&\frac{1}{2\pi v_F}{\lambda_2}^2+\frac{K}{2\pi
  v_F}{\lambda_3}^2\\ \nonumber
\frac{d\lambda_3}{dl}&=&\frac{1}{2}(1-K)\lambda_3+\frac{1}{\pi
  v_F}\lambda_2\lambda_3.
\end{eqnarray}

First, note that the first term in the third equation is seemingly different
from the one for the lattice case: it comes from the fact that the impurity
only acts at $x=0$, leaving only a double integral over time and
producing $\Delta_N=0$. The single-impurity gap (or 
the single-site Kondo temperature) now scales as
\begin{equation}
\label{gapi}
\Delta\propto {\lambda_3}^{\frac{2}{1-K}}\equiv 
{\lambda_3}^{\frac{2\pi v_F}{U}}.
\end{equation}
For $U\rightarrow 0$, marginal operators produce the usual Kondo temperature
[identical to Eq.(\ref{impurity})].
\vskip 0.1cm
The second difference compared to the finite density of impurities is the
{\it absence} of renormalization of the bulk exponent $K$, which clearly
displays that there is no source of Drude resistivity (nor scattering
effects) far away from the impurity spin.
Thus, the two $\beta$-functions (for $\lambda_2$ and $\lambda_3$, 
respectively)
become strongly coupled due to the presence of the extra term ${\cal 
O}({\lambda_3}^2)$ in the second equation. 
To show the existence of such 
a term, one can for instance utilize current algebera techniques
for the operators\cite{Ludwig}. 
\vskip 0.1cm
The charge field satisfies the operator
product expansion (OPE) of a U(1) Gaussian
model,
\begin{equation}
\label{1}
:\exp(i\sqrt{2\pi}\Phi_c(\tau))::\exp(-i\sqrt{2\pi}\Phi_c(0)):\ 
\propto\frac{1}{u\left|\tau\right|^K}\cdot
\end{equation}
One can also use the OPE of spinon pairs:
\begin{equation}
\label{2}
\hbox{Tr}[\Phi^{(1/2)}(\tau){\sigma}^a]
\cdot\hbox{Tr}[\Phi^{(1/2)}(0){\sigma}^b]\propto
i\epsilon^{abc}[J_L^c+J_R^c].
\end{equation}
All these operators act on $x=0$, and $\tau$ is close to 0.
The local impurity spin obeys a usual SU(2) Lie algebra:
\begin{equation}
\label{3}
[S^a,S^b]=i\epsilon^{abc}S^c.
\end{equation}
Exploiting Eqs.(\ref{1}),(\ref{2}) and (\ref{3})
and the fact that $\left|\tau\right|^K\sim dl$, one
inevitably finds the extra term in the second equation. We have also
applied the equality $uK=v_F$, from the LL theory in
presence of Galilean invariance (Appendix A).
Let
us insist on the fact that in the KLM at half-filling
the exponent $K^*$ tends to zero at
low temperatures so that
forward and backward Kondo exchanges are independent
also in the infra-red region.
It is not the case in the single impurity case;
the low-temperature behavior is well controlled by
$\lambda_2=\lambda_3\rightarrow +\infty$. 

\subsection{Transport and fixed points}

At high temperatures, the only effect of the interaction $\lambda_3$ is
therefore to change the Born amplitude of magnetic scattering.
The conductance is given by the effective scattering at
the scale $L=\exp l\sim 1/T$. Integrating out (\ref{ali}), one gets
a Landauer-type conductance:
\begin{equation}
G_0-G(T)=-\delta G(T)\propto {\cal R}(T)=D_m^{o}T^{\tilde{\mu}-2},
\end{equation}
with:
\begin{equation}
\tilde{\mu}=K+1. 
\end{equation}
$G_0=2e^2K/h$ 
is the conductance of the pure wire 
(again, without any reservoir-effect). Applying a generalized
Fermi Golden's rule at $T\neq 0$, one can check that
${\cal R}(T)\sim D_m^{o}(T)={\lambda_3(T)}^2$ is the resistance
 produced by a sole
impurity. It obeys:
\begin{equation}
\label{single-disorder-flow}
\frac{dD_m^o}{dl}=(2-\tilde{\mu})D_m^o.
\end{equation}
This produces the same thermal laws as a 
pointlike nonmagnetic
defect at $x=0$ (see, for instance\cite{Helene} and references
therein), i.e.
\begin{equation}
G_0-G(T)\propto T^{K-1}.
\end{equation}
\vskip 0.2cm
On the one hand, attractive interactions $(K>1)$ favor the formation of
singlet superconducting pairs that can completely
suppress the back flow at zero 
temperature. Here, the impurity becomes unscreened 
for ``vanishing'' attractive
interactions or at least as soon as $U>J_K$. In
comparison, remember that 
the {\it coherence} between localized spins in the lattice
problem
should prevent a superconducting-like ground state until attractive enough
electron-electron interactions, i.e. $K\geq K_c=3-2\Delta_N$. Furthermore, the
Drude weight or superfluid stiffness is already large
in the weak attractive patch $(\sim 2v_F)$.

One the other hand, if the backward and forward Kondo couplings are relevant
i.e. for $K<1$,
the weak coupling renormalization
group scheme ceases to be valid as soon as the Born amplitude 
${\lambda_3}^2$ is of order one (i.e. for
$T\sim\Delta$).  From
Eq.(\ref{gapi}) one can notice that the single-site Kondo temperature is
substantially enhanced by the strong repulsive interactions in the 1D
LL: Again, repulsive interactions promote SDW fluctuations in the electron
liquid that interact prominently with the localized impurity.
 The screening of the
impurity becomes prominent and we have the formation
of a strong nonmagnetic barrier at $x=0$. The fact that $\lambda_3$ flows
to strong couplings produces a new boundary condition at the origin:
\begin{equation}
<\Phi_c(0)>\ =\sqrt{\frac{\pi}{2}}\ \leftrightarrow\  
<\chi^{\dag}_-(0)\chi_+(0)>\ =\sqrt{\Delta},
\end{equation}
that pins the charge
cosine potential. The impurity is screened by a physical electron
and then holons now escape away from the impurity site (see
Appendix B): no kink
in $\Phi_c$ is then possible at the origin. By symmetry, two
different fixed points are authorized. 

First, one can consider weak
tunneling of electrons
through the impurity site. This fixed point has been analyzed in
refs.\cite{Furusaki,KLH} 
and still gives a power law for the conductance, but
with a different exponent than in the weak coupling limit:
\begin{equation}
\label{condu}
G(T)\propto T^{1/K-1}.
\end{equation}
Of course, it vanishes at zero temperature. The exponent $\tilde{\mu}$
is renormalized as:
\begin{equation}
\tilde{\mu}=1+1/K.
\end{equation}
In such case, backscattering off magnetic impurity
will tend to reduce considerably the current in the wire. From this point
of view, we obtain a perfect insulator at $T=0$: 
However, in the single impurity case
the d.c. conductivity still grows at low temperatures 
[see Eqs.(34),(\ref{condu})]. Since $K$ does not
change with $l$, the transition between zero/finite conductance here occurs
at $\tilde{\mu}=2$, i.e. in the neighborood of the noninteracting fixed
point $K_c=1$. This situation would require that the screening characteristic
time is very large compared to the tunneling one.

Second, one could rather investigate the opposite limit [i.e. screening
time $\ll$ tunneling time]. 
By extension of the noninteracting case, one could then expect
a nonperturbative ground state of Fermi {\it liquid} type. 
Using boundary conformal
field theory, we have proved that electron-electron interactions could
also produce
another ground state with typically
the same thermodynamic properties as a 1D Fermi
liquid, but with ``magnon'' spin quasiparticles and 
a nonuniversal Wilson ratio\cite{KLH}.

Numerical works are up to now not very numerous:
A recent DMRG study\cite{Wang}
and a quantum Monte-Carlo procedure\cite{Komnik} have both
checked that the impurity is well screened at low temperatures
and found for very {\it large} Hubbard interactions,
a susceptibility law $\chi(T)\sim 1/\Delta$, typical
of a heavy Fermi liquid. In such
limit, charge degrees of freedom are completely frozen and the
unusual power law behavior predicted in ref.\cite{Furusaki,KLH} cannot
arise anymore. This scenario maybe
could help in describing
the recent heavy-fermion behavior reported in the strongly correlated
(two-dimensional) material $Nd_{2-x}Ce_xCuO_4$\cite{Brugger}.
Furthermore, computing the susceptibility curve
at low temperatures, the authors of ref.\cite{Komnik} have reported a
non-Fermi liquid behavior with an anomalous exponent 
$\eta=1/K$\cite{Furusaki,remark4} for {\it weak} 
interactions, and a Fermi-liquid
exponent $\eta=2$ for {\it vanishing} interactions.

The fact that a simultaneous existence
of two leading irrelevant operators, one describing
Fermi-liquid behavior and the other describing Furusaki-Nagaosa anomalous
behavior [implying: screening time $\simeq$ tunneling time], should not be possible for weak interactions would deserve 
further numerical works.

A summary of the various equations and transport properties for the Kondo
lattice(s) and the single impurity case
is given in table 1.

\subsection{Array of independent impurities}

To achieve this part, let us briefly compare with the case
of $N_i={\cal L}/a$ {\it independent} magnetic impurities in
a wire of length $L={\cal L}$ (i.e.
the density of magnetic impurities is fixed to $n_i=1$). 
The total resistance
of the wire reads
\begin{equation}
{\cal R}_{tot}(T)=\sum_{N_i}\ D_m^{o}(T)=
\frac{{\cal L}}{a}D_m^{o}(T)\neq D_m(T).
\end{equation}
Now, using the precious link between conductance and conductivity in 1D
for a metallic system of size ${\cal L}$,
\begin{equation}
\label{uni}
G_{tot}={\cal L}^{-1}\sigma_{dc},
\end{equation}
one finds the high-temperature law:
\begin{equation}
\sigma_{dc}(T)\propto {\cal L}{\cal R}_{tot}^{-1}=a[D_m^{o}(T)]^{-1}\propto
\frac{1}{D_m^{o}}T^{1-K}.
\end{equation}
In the preceding studies of the Kondo lattices, we have obtained 
the same formula
replacing $\tilde{\mu}$ by $\mu$. 
Since $\mu\ll\tilde{\mu}$ [because
$\Delta_N\leq 1/2$ and especially because $K(T)$ becomes smaller and smaller
decreasing the temperature in the real
lattice problem], this leads to a {\it faster} decay of
conductivity/conductance
when the temperature decreases, compared to that of independent magnetic
impurities. 
Coherent effects between magnetic impurities are included
through the scaling dimension parameter
$\Delta_N$, and through the quantum
renormalization of the exponent $K$, as well. These definitely enhance Kondo
localization. Likewise, 
note that when the impurities are supposed to be completely
independent, localization should not occur for attractive interactions [see
Eq.(\ref{single-disorder-flow})].

\section{Conclusion}

To summarize, 
we have studied (two) metal-Kondo insulating transitions
possibly occurring in one dimensional quantum wires. These
result from a consequent antiferromagnetic (Kondo)
coupling between conduction
electrons very close to half-filling and a perfect lattice of local moments
--- producing a usual KLM. 

As a result of commensurability, we have shown how a
one-dimensional Kondo insulator can be understood as an effective (strong)
umklapp process becoming relevant. Therefore, studying the pure KLM 
in the strong Kondo coupling limit $J_K\gg t$, results in a
commensurate-incommensurate
transition of Pokrovsky-Talapov type at zero temperature. 
For instance, the charge
susceptibility should diverge as the inverse of the doping near half-filling,
that is in good agreement with recent numerical datas\cite{Shibata}. 

In the weak coupling region $(J_K,U)\ll t$, Kondo transitions can be
classified as a function of the coherence-parameter in the spin array i.e.
the scaling dimension of the staggered magnetization operator,
namely $\Delta_N$, which is always smaller than 1/2. These
arise due to the important renormalization
of the backward Kondo 
potential. In such regime, the interplay between electron-electron 
interactions and backward Kondo scattering is
predicted to produce exotic and fascinating features in transport properties,
in the high-temperature (ballistic) regime --- whatever $\Delta_N\leq 1/2$.

For repulsive interactions, the conductivity is found to decrease 
monotonically with temperature --- even
at high temperature: There is no remnant of the original umklapp process
driven by the on-site Hubbard interaction.
The resulting Kondo insulating phases are still stable in presence of weak
non-magnetic disorder. {\it These are 
perfect signatures of Kondo localizations}. 

In contrast, for weak 
attractive interactions, the electron liquid is less
localized by magnetic impurities, and then the d.c. conductivity yields a 
maximum in the entrance of the
localized phase. This is a precursor of the s-wave superconducting phase 
occurring for stronger attractive interactions. 

The Kondo localization should even persist when the impurities are supposed
to be randomly distributed. In particular, repulsive interactions
also help in reducing the conductivity stiffness at
intermediate length scales. As a consequence, persistent
currents on thin mesoscopic rings with prominent magnetic defects should
then \emph{decrease} for repulsive interactions.

Finally, for completeness, consequences of the
backward Kondo scattering on transport of LL's have been summarized
in Table 1, both in the one impurity case and that of a perfect lattice of
magnetic impurities. 

\acknowledgements{I thank T. Giamarchi and T. Maurice Rice
for relevant comments throughout this work. I also benefit from discussions
with P. Prelov\v{s}ek and A. Rosch on transport in low-dimensional systems.}

\appendix
\section{Duality spinless fermion-boson in one dimension}

We use a continuum
version of the Jordan-Wigner scheme to rewrite
the right and left electron fields $\Psi_q(x)$ (with $q=\pm$)
in a bosonic basis. We define
\begin{equation}
\Psi(x)=\Psi_+(x)+\Psi_-(x).
\end{equation}
and,
\begin{equation}
\psi_j\rightarrow \Psi(x)\sqrt{a}
\end{equation}
$a$ being the lattice spacing.
Each
fermion operator can be re-expressed as,
\begin{equation}
\Psi_q(x)=\frac{1}{\sqrt{2}}{\cal O}_q(x){\cal B}(x)=
\frac{1}{\sqrt{2}}\exp[i\pi q\sum_{y<x} n(y)]{\cal B}(x),
\end{equation}
where $n(x)=\sum_q\ \Psi_q^{\dag}(x)\Psi_q(x)$ is the total number operator, 
and ${\cal B}(x)$ is in principle
an explicit hard core boson operator. The extra
factor $1/\sqrt{2}$ allows
to fullfill the Jordan-Wigner constraint,
\begin{equation}
n(x)={\cal B}^{\dag}(x){\cal B}(x).
\end{equation}
By
construction of spinless fermions, we have: ${\cal B}^2(x)=0$. 
On the
other hand, as long as $\left|x-y\right|\geq 1$
(1 being now the lattice spacing), we can 
check that Bose operators always commute.
The trick is then to come back to a {\it free} boson system using the argument
that in very low density there is no real
difference between free- and hard
core bosons. To perform this explicitly, we rewrite
\begin{equation}
n(x)=\rho_o+b^{\dag}(x)b(x),\ \rho_o=\frac{1}{2}=\frac{k_F}{\pi}\cdot
\end{equation}
Considering a gas of spinless fermions near half-filling $[n(x)\rightarrow
1/2]$ results in:
\begin{equation}
\frac{\delta\rho}{\pi}=b^{\dag}(x)b(x)\rightarrow 0,
\end{equation}
and approximately,
\begin{equation}
b(x)=(\frac{\delta\rho}{\pi})^{1/2}\exp i\sqrt{\pi}\Theta_c.
\end{equation}
$\delta\rho$ measures density fluctuations and $\Theta_c$ is the
associated superfluid phase. This corresponds to the usual phase-amplitude
decomposition. It is convenient to introduce a phonon-like 
displacement field:
\begin{equation}
\delta\rho(x)=\sqrt{\pi}\partial_x\Phi_c(x).
\end{equation}
As usual, the density and phase are canonically conjugate quantum variables
taken to satisfy:
\begin{equation}
[\Theta_c(x),\partial_y\Phi_c(y)]=i\delta(x-y).
\end{equation}

This choice of wave-function is particularly judicious because
it satisfies the (last) condition that $\Psi_q(x)$ must 
remove a unit charge e at the coordinate x. To see this, one can rewrite:
\begin{equation}
e^{i\sqrt{\pi}\Theta_c}=e^{i\sqrt{\pi}\int_{-\infty}^{x}\ dx'\ 
\Pi_c(x')},
\end{equation}
and 
\begin{equation}
\Pi_c=\partial_x\Theta_c,
\end{equation}
is the momentum conjugate to $\Phi_c$. Since
$\Pi_c$ 
is the generator of translations in $\Phi_c$, this creates a kink in 
$\Phi_c$ of height 
$\sqrt{\pi}$ centered at $x$, which corresponds to a localized
unit of charge using the definition of $b^{\dag}(x)b(x)$.
Moreover, the Jordan-Wigner string can be rewritten:
\begin{eqnarray}
{\cal O}_q&=&\frac{1}{\sqrt{2}}\exp iq\int^x\ dx'\ (\delta\rho(x')+k_F)\\
\nonumber
&=&\frac{1}{\sqrt{2}}\exp iq(k_Fx+\sqrt{\pi}\Phi_c(x)).
\end{eqnarray}
We have thereby identified the correct Bosonized form for the
(continuum) electron operators:
\begin{equation}
\Psi(x)=\psi_{R}(x)e^{ik_F x}+\psi_{L}(x)e^{-ik_F x},
\end{equation}
with:
\begin{equation}
\psi_q=\frac{\eta_q}{\sqrt{{2}\pi a}}:\exp i\sqrt{\pi}(\Theta_c+q\Phi_c):.
\end{equation}
We have replaced the lattice step $a$ taken to satisfy Eq.(A3).
This construction has been built originally by Haldane. The expression
(A13) is called
normal ordered (: :) meaning that all $\infty$ constants have been
substracted from the fermionic ground state. In particular, we have:
\begin{equation}
\sum_q \psi_q^{\dag}(x)\psi_q(x)\propto\partial_x\Phi_c(x)\rightarrow 0.
\end{equation} 
We have added Klein factors defined by ${\eta_-}^2={\eta_+}^2=1$ and
${\eta_+}{\eta_-}+{\eta_-}{\eta_+}=0$, to perfectly
fullfill anticommutation rules between left and right movers.
Moreover, the density fluctuation field
$\partial_x\Phi_c$ creates propagating particles. This is
just a consequence of the fact
that in 1D electron-hole pairs are exactly {\it coherent} because they
propagate with the same velocity. The bosonization scheme just reflects
this important point.

Then, one can easily prove
that the so-called Dirac Hamiltonian $H=\int dx\ {\cal H}$ with
Hamiltonian density
\begin{equation}
{\cal H}=-iv_F[\psi_+^{\dag}\partial_x \psi_+ - 
\psi_-^{\dag}\partial_x \psi_-],
\end{equation}
can be exactly rewritten as a wave propagating at velocity $u=v_F$
(and $K=1$):
\begin{equation}
{\cal H}=\frac{u}{2\pi}[\frac{1}{K}:(\partial_x\Phi_c)^2:+K:(\Pi_c)^2:].
\end{equation}
Remarkably, adding a {\it weak}
Hubbard interaction between electrons results {\it only} in a renormalization
of $K$ and $u$ as
\begin{equation}
uK=v_F\ \hbox{and}\ K=1-\frac{Ua}{\pi v_F}\cdot
\end{equation}
We obtain the general Luttinger liquid Hamiltonian.

\section{Weak coupling and fermionic kinks}

In the weak coupling limit $J_K/t\ll 1$ and $U/t\ll 1$ and at 
half-filling, one must start with
$K\rightarrow 1$. The density of holons
at $q=4k_F\sim 0$ is described by\cite{KV}:
\begin{equation}
\chi_-^{\dag}(x)\chi_+(x)\propto \exp[i\sqrt{2\pi}\Phi_c(x)].
\end{equation}
This has a scaling dimension close to 1/2. We immediately deduce
that the scaling dimension of the holon field is 1/4. Then, for the
electron {\it gas}, the 
holon (like the spinon, see Appendix C) behaves 
as a {\it semion} or half of an electron\cite{BS}. This is equivalent to say
that the free electron decomposes itself into spinon and holon.
\vskip 0.1cm
At low temperatures and half-filling, holons 
acquire a mass due to the strong renormalization of the
backward Kondo exchange, and charge carriers rather behave as
{\it fermionic kinks}: {\it The Kondo coupling is
equivalent to an umklapp process}. To show that in a quite exact
manner, we adopt
the following scheme.
First, it is convenient to perform an average on the spin sector:
\begin{equation}
<\hbox{Tr}\vec{\sigma}\Phi^{(1/2)}\cdot\vec{N}>\ =
m^{\frac{\lambda^2}{4\pi}}
:\hbox{Tr}\vec{\sigma}\Phi^{(1/2)}\cdot\vec{N}:,
\end{equation}
with $m=\Delta$ (the spin gap at the Fermi level) and:
\begin{equation}
{\lambda}^2=2\pi(1+2\Delta_N).
\end{equation}
At the crossover temperature, we have:
\begin{equation}
:\hbox{Tr}\vec{\sigma}\Phi^{(1/2)}\cdot\vec{N}:\ =\hbox{constant}.
\end{equation}
The Kondo exchange turns the spinons of the 1D LL into optical magnons.
Using the gap-equation (\ref{gap}), this results in
\begin{equation}
J_K<\hbox{Tr}\vec{\sigma}\Phi^{(1/2)}\cdot
\vec{N}>\ \propto \Delta^{2-\frac{K}{2}}.
\end{equation}
The Kondo interaction then reads:
\begin{equation}
{\cal H}_{hf,c}=\frac{\Delta^{2-\frac{K}{2}}}{2\pi a}
\cos\sqrt{2\pi K}\Phi_c.
\end{equation}
Second, when the bare LLP is (nearly) 1, one gets:
\begin{equation}
{\cal H}_{hf,c}=\frac{\Delta^{3/2}}{2\pi a}\cos\sqrt{2\pi}\Phi_c
=\frac{\Delta}{2\pi a}\cos\sqrt{4\pi}\Phi_c.
\end{equation}
Using the definition (\ref{sept}), one definitely obtains:
\begin{equation}
{\cal H}_{hf,c}=-i\Delta(\psi_{+}^{\dag}\psi_{-}-\psi_{-}^{\dag}\psi_{+}),
\end{equation}
that typically leads to the same spectrum given in Eq.(\ref{mass}). Again, we
have chosen Klein factors such as $\eta_+\eta_-=+i$. 
As soon as $J_K\neq 0$, we obtain a Kondo insulator:
Pairs of holons cannot be excited at zero temperature and typically the LLP
tends to {\it zero} (exactly at half-filling). Note that other
excitation branches (`breathers'), which correspond to kink-antikink bound
states, cannot appear because the effective dimension of the
SG-operator is 1. 
\vskip 0.25cm
\emph{The backward Kondo coupling is then
equivalent to an umklapp process with $K=1/2$ and $g_{3\perp}=\Delta$ at
low temperatures and low freqencies.} However, conversely to the strong
coupling limit $J_K/t\gg 1$, this picture is here only
appropriate at low energy. 
Charge excitations at high temperatures are rather semions. This produces
differences e.g. in transport features at high frequency.

\section{Field theories for spin degrees of freedom}

Considering low-dimensional
spin problems with SU(2) symmetry, we obtain a particular class of 
conformally invariant theories
which has an Hamiltonian density quadratic in the currents 
$J^a(x)$ $(a=x,y,z)$\cite{Affleck_boso}: 
\begin{equation}
\label{h}
{\cal H}_s(x)=\frac{1}{2+k}: {\vec J}(x)\cdot {\vec J}(x):.
\end{equation}
The velocity of spin excitations has been fixed to 1.
Here $k$ is the Kac-Moody level which must be a positive integer.
It follows that the 
Sugawara density Hamiltonian obeys the so-called Virasoro algebra with a
conformal anomaly parameter: $C=3k/(2+k)$. 

\subsection{Heisenberg model}

The low-temperature behavior of a single Heisenberg chain is 
described by a Sugawara Hamiltonian with $k=1$.
  
The physical particles (pairs of spin-1/2 excitations or spinons),
are included through the primary 
fields $\Phi^{(1/2)}$ and $\Phi^{(1/2)\dag}$ from the representation of the 
$SU(2)$ group. The action taking into account 
the dynamics of the spinon pairs is given by:
\begin{eqnarray}
S_{WZW}&=&-\frac{1}{16\pi}\int d^2 x\ 
\hbox{Tr}\hbox{\large(}
\partial_{\mu}\Phi^{(1/2)\dag}\partial_{\mu}\Phi^{(1/2)}\hbox{\large)}
\\ \nonumber
&+&\frac{i}{24\pi}\int_0^{\infty} d\epsilon\int d^2 x
\epsilon^{\alpha\beta\gamma} \hbox{Tr}
\hbox{\large(}{\cal A}_{\alpha}{\cal A}_{\beta}{\cal A}_
{\gamma}\hbox{\large)},\\ \nonumber
\end{eqnarray}
and, 
\begin{equation}
{\cal A}_{\alpha}=\Phi^{(1/2)\dag}\partial_{\alpha}\Phi^{(1/2)}.
\end{equation}
The last term is from topological origin. The $2k_F$ SDW operator 
(i.e. the $2k_F$ spinon density) can be
identified as:
\begin{equation}
\vec{N}(x)=e^{i2k_Fx}\hbox{Tr}(\Phi^{(1/2)}+\hbox{Tr}\Phi^{(1/2)\dag})\vec{\sigma}.
\end{equation}
and has a scaling dimension 1/2. We immediately deduce that a single
spinon at $k=k_F$ has a scaling dimension 1/4 and behaves as a semion
\cite{BS,Haldane1}.

\subsection{Usual Spin ladder}

Here, it is natural to rewrite the theory in terms of the total spin current:
\begin{equation}
\vec{R}_{L/R}(x)=\vec{J}_{L/R}(x)+\vec{I}_{L/R}(x).
\end{equation}
These obey the Kac-Moody algebra with $k=2$. The associated Hamiltonian,
constructed from $\vec{R}(x)$, ${\cal H}_s$, has C=3/2. 
A crucial point is that ${\cal H}_{sJ}+{\cal H}_{sI}$ can be written as a sum
of two commuting pieces, ${\cal H}_s$ and a remainder. The value of the
conformal anomaly for this remainder Hamiltonian is $C=2-3/2=1/2$. There is
a unique unitary conformal theory
 with this value of C, namely the {\it Ising} model:
This is an
example of Goddard-Kent-Olive coset construction\cite{GKO}.
The eigenstates in the $SU(2)_{2L}\times SU(2)_{2R}$ sector appear in
conformal towers labeled by spin quantum numbers $j=0,1/2,1$. The
corresponding primary fields are: the identity {\bf 1}, the fundamental field
${\gamma}^{(1/2)}$, and the triplet operator (a $3\times 3$ matrix) 
${\bf \Phi^{(1)}}=\sum_{i,j}\Phi_{Lj}\Phi_{Rj}$. Their scaling dimension
is given by $0,3/8,1$ respectively.
Similarly, there are three primary fields in the Ising sector: the identity
operator ${\bf 1}$, the Ising order parameter $\sigma$ and the energy operator
$\epsilon$ with scaling dimensions $0,1/8,1$ respectively. 

The spin ladder system is then described by the effective 
Hamiltonian\cite{Totsuka-Suzuki}:
\begin{equation}
\label{C13}
H=H_s^{(k=2)}+H^{(Ising)}
+\int dx\ [\Delta\hbox{Tr}{\bf\Phi^{(1)}}-3\Delta\epsilon].
\end{equation}
$\Delta$ is typically the interchain coupling. 

Notice that starting with two
different intrachain Heisenberg coupling constants produces an extra term in
the action, namely:
\begin{equation}
\delta S\propto
\frac{|J_{\parallel 1}-J_{\parallel 2}|}{max(J_{\parallel 1},
J_{\parallel 2})} \int
d^2x\ \hbox{Tr}\hbox{\large(}
\partial_{\mu}\Phi^{(1/2)\dag}\partial_{\mu}\Phi^{(1/2)}\hbox{\large)}.
\end{equation}
Redefining $\Phi^{(1/2)}=i\vec{\sigma}\cdot{\bf\Phi}$, the result is an O(3)
nonlinear sigma model built out of the field ${\bf\Phi}$. Since the
topological term has here no contribution, then the gapless ordered
state of the isotropic sigma model is unstable. The spin gap should be
rescaled as:
\begin{equation}
\Delta_S\propto \Delta 
\exp-\hbox{\Huge(}\frac{|J_{\parallel 1}-J_{\parallel 2}|}{max(J_{\parallel 1},
J_{\parallel 2})}\hbox{\Huge)}\cdot
\end{equation}
The distinction between $\Delta$ and $\Delta_S$ can be done only for
appreciable differences between $J_{\parallel 1}$ and $J_{\parallel 2}$.

\subsection{Takhtajan-Babujan chain}

Now, we start with the integrable Takhtajan-Babujan S=1 chain on a lattice,
described by:
\begin{equation}
H=J_{\parallel}\sum_j ({\bf S}_j{\bf S}_{j+1})-
\beta({\bf S}_j{\bf S}_{j+1})^2,
\end{equation}
where $\beta=1$\cite{TB}. The unusual term quartic in spin 
can be generated, for 
example, by phonons. Solving Bethe Ansatz
equations, the only elementary excitation is known to be a doublet
of gapless spin-1/2 spin waves, with total spin 0 or 1. The specific
heat behaves as: $C_v/L=2S T/(1+S)+{\cal O}(T^3)$, where S=1. Then, in
the sense of critical theories, such a model could be parametrized by a 
conformal anomaly $C=3S/(S+1)=3/2$.
This fact, together with numerical checks, leads to
the conclusion that the criticality of this model is governed by operators
satisfying a 
critical Sugawara model with 
$k=2S=2$. In particular, the staggered
magnetization $\vec{N}$ can be simply 
expressed as:
\begin{equation}
\vec{N}(x)=
\hbox{Tr}\{({\gamma}^{(1/2)}+{\gamma}^{(1/2)\dag})\vec{\sigma}\},
\end{equation}
with scaling dimension 3/8. 
The Hamiltonian is in turn equivalent to three Majorana fermions (or
triplet excitations, and not doublets).

\section{Links with non-magnetic Gaussian disorder}

Let us now make comparisons with transport properties of a LL with many
randomly distributed {\it non-magnetic} impurities. 

For not too strong disorder, one usually approximates
the disorder by a random potential,
\begin{equation}
H_{dis}=\int dx\ V(x)\rho^{(2k_F)}(x).
\end{equation}
$\rho^{(2k_F)}$ is the $2k_F$ charge density operator.
As usual, we can omit forward scattering because it can be incorporated
into the shift of the chemical potential and does not affect the fixed point
properties. $V(x)$ is generally Gaussian correlated i.e.:
\begin{equation}
\overline{V(x)}=0,\qquad \overline{V^*(x)V(x')}=D\delta(x-x').
\end{equation}
One also assumes that the concentration of impurities becomes
infinite $n_i\rightarrow +\infty$, but that the scattering potential of
each impurity becomes weak $V\rightarrow 0$ so that the product:
\begin{equation}
D=n_iV^2,
\end{equation}
remains a {\it constant}. Note the manifest
difference with the Kondo lattice at half-filling.
For $\Delta_N=0$ and high temperatures, 
the RKKY interaction produces a constant (flat)
uniaxial potential:
\begin{equation}
\overline{V(x)}=\lambda_3,\qquad \overline{V(x)V(0)}=D_m={\lambda_3}^2.
\end{equation}
Adding an explicit antiferromagnetic exchange between local moments creates
rather an isotropic potential; each component satisfies (i=x,y,z):
\begin{eqnarray}
\overline{V(x)}&=&\lambda_3<\hbox{Tr}\hbox{\large(}\Theta^{(1/2)}(x){\sigma}^i
\hbox{\large)}>\ =0,\\ 
\nonumber
\overline{V(x)V(0)}&=&D_m<N^i(x)N^i(0)>\ =D_m/x^{2\Delta_N}.
\end{eqnarray}
(The two-point correlation functions are here computed for equal times).
{\it For simplicity, in the following, we consider the bare
lattice step equal to
one}.
For spins S=1/2 and a resulting Heisenberg exchange, we have $\Delta_N=1/2$.
For an S=1 Takhtajan-Babujan chain, the result is $\Delta_N=3/8$. The
averages are essentially performed on {\it quantum} fluctuations. By analogy 
to non-magnetic disorder, we have defined:
\begin{equation}
H_{mag}=\int dx\ \vec{V}(x)\cdot\vec{\rho_s}^{(2k_F)}(x),
\end{equation}
with:
\begin{equation}
V_i(x)=V(x)=\lambda_3 N^i(x).
\end{equation}
In KLM's with $n_i=1$, the magnetic potential has a 
Fourier component at $q=\pi$, and then it is only relevant at half-filling.

In the interplay between (Gaussian correlated)
nonmagnetic disorder and
interactions, the memory function approximation
does not give the correct result. An extra term occurs coupling disorder
and spin backscattering. We must proceed as follows.

In the pure system, one has to include the spin interaction:
\begin{equation}
{\cal H}_{bs}=\frac{2g_{1\perp}}{(2\pi)^2}\int dx\ 
\cos\sqrt{8\pi}\Phi_s(x).
\end{equation}
written in the Abelian formalism\cite{Helene}. For
small U, the spin Luttinger parameter reads
\begin{equation}
K_s=1+\frac{g_{1\perp}}{\pi v_F}=1+\frac{U}{\pi v_F}\cdot
\end{equation}
For repulsive interactions, the fixed point is described by $K_s^*=1$
and $g_{1\perp}^*\rightarrow 0$. For attractive interactions, $g_{1\perp}$
flows to $-\infty$ and opens a superconducting gap. But, in both
cases, $g_{1\perp}$ affects much transport properties at intermediate length
scales. In the Born (or still Hartree-Fock) 
approximation, one can write\cite{Helene}:
\begin{equation}
\sigma_{dc}(T)=T^{-1}/D(T)=l_e(T).
\end{equation}
The conductivity, that is a physical quantity, is of course not changed 
under renormalization.
The (dimensionless) non-magnetic disorder obeys:
\begin{equation}
D=l_{in}/l_e,
\end{equation}
where $l_e$ is the ``effective'' 
elastic mean-free path and $l_{in}=v_F/T$, the
thermal length at which renormalization procedure must stop due to
the generation of {\it inelastic} processes. 
We make the approximation that the
equation flows are not modified up to this length scale. The obtained
dc-conductivity corresponds to the one of the infinite system.

The delocalized regime 
is inevitably driven by
quantum effects. The Anderson- (or still Kondo)
localization length $\L_{loc}\sim \Delta^{-1}$ 
is very short in 1D: {\it There is no diffusive or
Boltzmann region, and then
the mean-free path is not a universal quantity 
(the Einstein diffusion constant)}. 
\vskip 0.2cm
Using the recursion law found in ref.\cite{Giam2}:
\begin{equation}
\frac{dD}{d\ln T}=-(3-K_s(T)-K-\frac{g_{1\perp}}{\pi v_F})D,
\end{equation}
one gets:
\begin{equation}
\sigma_{dc}(T)=T^{2-\hat{\mu}}\qquad \hat{\mu}\rightarrow 
K_s(T)+K+\frac{g_{1\perp}}{\pi v_F}\cdot
\end{equation}
This is the {\it precise} exponent of the d.c. conductivity at finite
length scales. For small $U$, one obtains:
\begin{equation}
\sigma_{dc}(T)\sim T^{-U/\pi v_F}.
\end{equation}
{\it Conversely to the case of magnetic impurities, 
repulsive interactions tend to make the disorder less relevant
and decrease localization}.

For sufficiently strong attractive interactions, one gets:
\begin{equation}
\sigma_{dc}(T)=T^{2-K}.
\end{equation}
(One has to exclude the term in $g_{1\perp}D$ because it becomes
 nonperturbative:
its effect is anyway to create a gap resulting in $K_s^*\rightarrow 0$
and a quite small localization length).

For the 1D KLM's with $n_i=1$, one can also write [from
Eq.(\ref{reclaw})]:
\begin{equation}
\frac{dD_m}{d\ln T}=-(3-\mu)D_m
\end{equation}
so that:
\begin{equation}
\sigma_{dc}(T)=T^{-1}/D_m(T).
\end{equation}
This implies that at high temperatures
we have a perfect duality to a model, where the distribution
of the magnetic potential is
rather Gaussian correlated and the effective
exponent for the
spin density-spin density correlation function in
the electron liquid is $\mu=K(T)+2\Delta_N$.
It becomes then simple to investigate the competition between Anderson- and
Kondo localization(s). For small $U$, one gets:
\begin{eqnarray}
\mu &=& 1-\frac{U}{\pi v_F}+2\Delta_N\leq 2-\frac{U}{\pi v_F},\\ \nonumber
\hat{\mu} &=& 2+\frac{U}{\pi v_F}\cdot
\end{eqnarray}
For repulsive interactions, one finds $\mu<\hat{\mu}$, producing Kondo
localization whereas attractive interactions favor the Anderson glass
to arise.

\section{Third order terms for Gaussian magnetic randomness}

We follow exactly the same scheme as in Appendix A of ref.\cite{Giam2}.
At lowest order, like for non-magnetic randomness, one obtains:
\begin{eqnarray}
\frac{dD_m}{d\ln L} &=& (3-K_s-K)D_m,\\ \nonumber
\frac{dg_{1\perp}}{d\ln L} &=& 2(1-K_s)g_{1\perp}.
\end{eqnarray}
Here, $D_m$ and $g_{1\perp}$ are {\it dimensionless} parameters.
On the other hand,  a third order term is also involved in the computation
of 2-point correlation function ${R}_{\nu}$ $(\nu=s,c)$, which is:
\begin{eqnarray}
& &{R}_{\nu}^{\hbox{(III)}}(r_1-r_2)=-\hbox{const.}\frac{g_{1\perp}}{a^2}\frac{D_m}{a^2}\sum_{\epsilon_6=\pm}\int
[...]\\ \nonumber
&\times & <T_{\tau}\ e^{i\sqrt{2\pi}\Phi_{\nu}(r_1)}
e^{-i\sqrt{2\pi}\Phi_{\nu}(r_2)}\cos\sqrt{8\pi}\Phi_s(r_3)\\ \nonumber
&\times & \sin\sqrt{2\pi}\Phi_s(r_4)\sin\sqrt{2\pi}\Phi_s(r_5)e^{i\epsilon_6
[\sqrt{2\pi}\Phi_{c}(r_4)-\sqrt{2\pi}\Phi_{c}(r_5)]}>
\end{eqnarray}
with,
\begin{equation}
[...]= dx_3 d\tau_3 dx_4 d\tau_4 dx_5 d\tau_5\delta(x_4-x_5).
\end{equation}
If the points $(x_4,\tau_4)$ and $(x_5,\tau_5)$ are close together in
a ring of inner radius a and width $da$, then the
element of integration becomes:
\begin{equation}
[...]=\hbox{const.}da\int\int\int\int dx_3 d\tau_3 dx_4 d\tau_4.
\end{equation}
When contracted and summed over $\epsilon_6$ the
\begin{equation}
<e^{i\epsilon_6
[\sqrt{2\pi}\Phi_{c}(r_4)-\sqrt{2\pi}\Phi_{c}(r_5)]}>=\exp{\hbox{\Large(}-F_c
[0,\frac{a}{v_F}]{\hbox{\Large)}}}
\end{equation}
part gives a constant factor (that is independent of the lattice cut-off), and
from the eliminated degrees of freedom we obtain:
\begin{eqnarray}
& & + \hbox{const.}\ da\frac{g_{1\perp}}{a^2}\frac{D_m}{a^2}\int\int\int\int dx_3d\tau_3 dx_4 d\tau_4 <T_{\tau} \\ \nonumber
& &e^{i\sqrt{2\pi}\Phi_{\nu}(r_1)}
e^{-i\sqrt{2\pi}\Phi_{\nu}(r_2)}\cos\sqrt{8\pi}\Phi_s(r_3)\cos\sqrt{8\pi}\Phi_s(r_4)>
\end{eqnarray}
which is {\it identical} to a $g_{1\perp}^2$ term and contributes only to
$R_s$. In particular, redefining $D_m\rightarrow aD_m$ (the dimensionless
disorder) one gets:
\begin{equation}
g_{1\perp}^2(l+dl)=g_{1\perp}^2(l)+\hbox{const}.\ dlD_mg_{1\perp}(l).
\end{equation}
As usual, one puts: $a=e^l$ and $\frac{da}{a}=dl$. The recursion law
for the spin backscattering becomes exactly:
\begin{equation}
\frac{dg_{1\perp}}{d\ln L} = 2(1-K_s)g_{1\perp}+D_m.
\end{equation}
\vskip 0.1cm
If the points $(x_3,\tau_3)$ and $(x_4,\tau_4)$ [or $(x_5,\tau_5)$] are
close together, one obtains an extra renormalization of $D_m$,
\begin{equation}
D_m(l+dl)=D_m(l)+\frac{g_{1\perp}}{\pi v_F}\ dl D_m(l),
\end{equation}
and finally:
\begin{equation}
\frac{dD_m}{d\ln L} = (3-K_s-K+\frac{g_{1\perp}}{\pi v_F})D_m.
\end{equation}
The sign of the last term is opposite to the one in the non-magnetic
disorder problem. This reveals a strong competition between Kondo localization
and superconductivity for attractive enough electron-electron interactions.

\onecolumn
\vskip 0.3cm
\begin{center}
\begin{tabular}{||c|c|c|c|c|c|c||}
\tableline
Impurities & $D_m$ & $dK/dl$ & $dD_m/dl$ & exponent & $\sigma_{dc}$ or 
$\delta G$  & 
transition\\
\tableline
lattice & ${\lambda_3}^2$ & $-{K}^2 D_m(l)$ & $[3-\mu(l)]D_m$ & 
$\mu=K(T)+2\Delta_N$ &
$\sigma \sim T^{2-\mu[T]}$ & Kondo insulator(s) \\
\tableline
single & ${\lambda_3}^2$ & 0 & $[2-\tilde{\mu}(l)]D_m$ & $\tilde{\mu}=K+1$ &
$\delta G \sim T^{\tilde{\mu}-2}$ &
Boundary effects \\
\tableline
\end{tabular}
\end{center}
\vskip 0.3cm
{\it Table 1: Weak coupling.---
Renormalization of the various parameters, physical properties
and transport properties both in the case of
Kondo lattices and of a single magnetic impurity. In this Table, for
simplicity we nominate $D_m$, the magnetic disorder parameters in both 
limits. In the case of randomly distributed magnetic defects, we have found
$\sigma_{dc}(T)\propto T^{Ua/\pi v_F}$ and a glassy phase which should
correspond to a disordered Ising phase.
For a comparison with 
non-magnetic impurities, see for instance ref.\cite{Helene}.}


\begin{references}
\bibitem{remark1}
A general scaling theory predicts: $D\propto \delta^{1+\frac{z-2}{d}}$, where
$d$ is the dimension of space.
\bibitem{Andrei-Rosch}
It is argued that to get such behavior, scattering processes between holons
and Kondo singlets is implicitly needed at finite temperatures. This
can be also provided e.g. by phonons, non-magnetic disorder.
\bibitem{remark4}
From scaling arguments, we obtain that the low-temperature
susceptibility obeys $\chi(T)=T^0+T^{2\delta_i-1}$ where $\delta_i$ is
the dimension of the given irrelevant operator. In the present case:
$2\delta_i=1+1/K$.
\end{references}
\end{document}